# AI Surrogate Modeling for Real-Time Tokamak Equilibrium Prediction: Benchmarking Neural Architectures and Validation on EXL-50U

Guoyang Shi[1,2,3], Zitong Zhang[4, 5], Siqi Ding[1,2,3], Jianguo Chen[1,2,3], Yapeng Zhang[1,2,3], Jiayi Zhi[6], Hanyue Zhao[1,2,3], Tianyuan Liu[1,2,3]*

*1 Beijing ENN Fusion Energy Science and Technology Co., Ltd., Beijing 101111, China*

*2 Beijing Key Laboratory of High Magnetic Field Spherical Torus Fusion Energy, Beijing 101111, China*

*3 Hebei Key Laboratory of Compact Fusion, Langfang 065000, China*

*4 College of Mechanical Engineering, Xi'an University of Science and Technology, Xi'an 710054, China;*

*5 MOE Key Laboratory of Thermo-Fluid Science and Engineering, School of Energy and Power Engineering, Xi'an Jiaotong University, Xi'an 710049, China*)

*6 Institute of Robotics and Automatic Information System, Nankai University, Tianjin 300350, China*

*Corresponding author: liutianyuan@enn.cn

**Abstract:** Fast and reliable plasma equilibrium prediction is essential for real-time tokamak operation, control, and fusion applications. However, conventional numerical solutions of the Grad-Shafranov (GS) equation are often too computationally expensive for real-time deployment. To address this challenge, we develop an AI surrogate framework and systematically evaluate five representative architectures (MLP, CNN, FNO, Transformer, and KAN) using a numerically generated GS equilibrium database comprising 100,000 IID samples and 10,000 OOD samples. Under a unified protocol, we assess prediction accuracy, inference efficiency, model scaling, and robustness under both Independent and Identically Distributed (IID) and Out-of-Distribution (OOD) conditions. We further establish a device-level validation chain on the EXL-50U tokamak by linking GS numerical solutions, surrogate predictions, and the standard Shape Editor solver reference, thereby explicitly evaluating simulation-to-device consistency. The surrogate models achieve errors of $10^{-3}$~$10^{-2}$ relative to GS solutions, while the discrepancy between GS solutions and the device reference remains at the $10^{-3}$ level. Transformer achieves the highest IID accuracy, whereas CNN provides a favorable balance between prediction accuracy, robustness, and inference efficiency, with a TensorRT latency of 0.7 ms. OOD experiments across unseen plasma geometries and parameter regimes reveal substantial architecture-dependent differences: CNN and FNO maintain superior extrapolation stability (relative $L_2$ error = 4%–5%), while architectures with weaker inductive biases exhibit greater performance degradation. Scaling experiments further show that increasing dataset size and model capacity consistently improves interpolation accuracy but does not necessarily enhance extrapolation, indicating a trade-off between model capacity and OOD generalization. Overall, this work establishes a systematic benchmarking and device-consistent validation framework for AI-based GS equilibrium prediction. The results demonstrate the potential of data-driven surrogates for real-time tokamak operation and provide practical guidance for selecting reliable architectures for future plasma control and fusion applications.

## 1. Introduction

Driven by global sustainability goals, magnetic confinement fusion, exemplified by tokamaks, has long been regarded as a leading pathway toward clean and virtually inexhaustible energy[1]. Recent studies increasingly indicate that artificial intelligence (AI) can play a transformative role in accelerating progress in magnetic confinement fusion. Representative applications include early warning of disruptive instabilities, real-time estimation of disruption proximity, and cross-device generalization through transfer learning, thereby casting disruption forecasting as a quantitative risk-prediction problem[2]. Control-oriented approaches can further estimate disruption proximity in real time, including the probability of occurrence and time to disruption, providing a deployable basis for disruption avoidance and mitigation in next-generation devices[3]. In addition, transfer-learning frameworks have demonstrated cross-device and cross-configuration disruption prediction, helping alleviate data scarcity when developing models for new machines[4]. Conversely, fusion research also provides a highly demanding testbed for scientific machine learning because of its multiphysics coupling, stiff nonlinear dynamics, sparse and noisy diagnostics, and strict safety constraints. These characteristics continue to drive advances in uncertainty quantification, interpretable modeling, and safe decision-making[5] [5]. Reflecting this bidirectional synergy, the AI-Fusion Digital Convergence Platform has been identified as a strategic priority in the U.S. Department of Energy’s 2025 Fusion Science and Technology Roadmap and as a key initiative supporting progress toward the commercialization of fusion energy.

Among the various magnetic confinement concepts, the tokamak remains one of the most technologically mature and is widely regarded as a leading candidate for realizing fusion power. The stable and high-performance operation of a tokamak critically depends on the accurate characterization and control of magnetohydrodynamic (MHD) equilibrium[6]. MHD equilibrium describes the force balance between the plasma pressure gradient and the Lorentz force, providing the theoretical foundation for numerous essential physics analyses, including MHD stability assessment, energy and particle transport modeling, and scenario optimization [7]. It also plays a central role in determining the operational limits of a tokamak. From an operational perspective, accurate real-time reconstruction of plasma equilibrium and control of the plasma position and shape are essential for maintaining stable operation, supporting disruption avoidance, and achieving advanced high-confinement regimes. Under the assumption of axisymmetry, the inherently three-dimensional equilibrium problem reduces to the two-dimensional nonlinear elliptic Grad-Shafranov (GS) equation[8]. Consequently, fast and accurate GS equilibrium computation constitutes a critical link among fusion theory, experimental control, and reactor design, particularly in real-time applications where discrepancies between simulation results and device-relevant conditions can directly affect control performance[9].

However, the development of GS solvers has long involved a practical trade-off between computational speed and solution accuracy, which becomes especially restrictive when transitioning from offline modeling to real-time experimental applications[10]. On the one hand, equilibrium computation within a real-time plasma control loop must be completed on millisecond or even sub-millisecond timescales to support control frequencies of up to several kilohertz[6]. On the other hand, in integrated modeling platforms that simulate complete plasma discharges, a GS solver may be invoked thousands of times, making its cumulative computational cost a limiting factor in scientific exploration and scenario optimization[11,12]. This challenge is expected to become even more pronounced for next-generation fusion devices such as ITER and DEMO, where both real-time control and large-scale integrated modeling impose increasingly stringent requirements on computational efficiency and solution fidelity.

Traditional GS methods address static or quasi-static MHD equilibrium, whereas fully time-dependent plasma evolution requires substantially more expensive simulations that are generally unsuitable for direct use in real-time contro[13,14]. Analytical equilibrium models provide rapid solutions but typically rely on simplified source terms. In contrast, numerical iterative solvers based on finite-difference or finite-element methods can accommodate nonlinear plasma profiles and therefore remain dominant in current practice[13]. From the perspective of the computational task, GS-related equilibrium calculations include forward equilibrium solving[15] and inverse equilibrium reconstruction from diagnostic measurements[10]. The latter seeks to infer plasma equilibrium from available diagnostic signals, with EFIT serving as a representative framework that incorporates external magnetic measurements under GS equilibrium constraints. Nevertheless, the iterative computational cost and sensitivity of convergence to initialization, profile parameterization, and numerical settings can still hinder high-bandwidth real-time applications and large-scale parameter scans.

In recent years, data-driven deep-learning methods[16] have emerged as a promising paradigm for alleviating efficiency bottlenecks in scientific computing. Their core idea is to use neural networks to learn complex input–output mappings from large-scale datasets generated by high-fidelity numerical solvers, thereby constructing accurate surrogate models capable of rapid inference. In magnetic confinement fusion, this approach has demonstrated substantial potential in disruption prediction[17], confinement-state identification[18], and plasma equilibrium reconstruction, indicating its ability to represent the strong nonlinearities inherent in fusion systems. Meanwhile, representative neural architectures, including convolutional neural networks (CNNs)[19–21], Fourier neural operators (FNOs)[22,23], Transformers[24] multilayer perceptrons (MLPs)[25], and Kolmogorov-Arnold networks (KANs)[26], have shown strong capabilities in approximating parametric partial differential equation solutions and nonlinear input–output mappings. These developments provide a methodological foundation for constructing fast scientific computing surrogates. Recent studies have further examined their effectiveness across equilibrium-related tasks, ranging from CNN-based plasma-boundary prediction[27] to magnetic-field and MHD-state prediction using FNOs and physics-informed neural operators (PINOs)[22,28,29], as well as the fusion of magnetic diagnostic data using Transformers[4] Collectively, these advances demonstrate the growing potential of learning-based models for addressing computational challenges in plasma equilibrium modeling.

Building on this foundation, increasing attention has been directed toward applying deep learning specifically to tokamak equilibrium computation, including equilibrium prediction and reconstruction, for which computational efficiency and real-time capability are critical. Lu et al. developed a fully connected neural network that uses magnetic measurements from the EAST tokamak to reconstruct the poloidal flux function, producing poloidal-flux estimates at a spatial resolution comparable to that of offline EFIT while satisfying the millisecond-level latency requirements of real-time plasma control. Joung et al. introduced GS-DeepNet, which incorporates the GS equation and external magnetic measurements into a fully connected neural framework. This approach enables unsupervised reconstruction of plasma equilibria formulated as inverse free-boundary problems without relying on EFIT-generated labels. Building on neural-operator methodologies, Bonotto et al.[30] incorporated the GS equation and boundary conditions into a convolutional physics-informed neural operator, demonstrating fast and accurate prediction of plasma equilibria and separatrix geometry. These studies collectively highlight the growing potential of physics-integrated deep-learning methods for efficient GS-related equilibrium computation.

Although previous studies have explored data-driven approaches to GS equilibrium computation, the reliability and practical deployability of these approaches remain insufficiently characterized in several important respects: 1) their ability to represent equilibria with complex and highly nonlinear current profiles; 2) their generalization from independent and identically distributed (IID) test samples to out-of-distribution (OOD) plasma geometries and parameter regimes; and 3) the availability of a unified framework for comprehensive and fair benchmarking across representative neural architectures. As summarized in Table 1, most existing studies focus on individual architectural or methodological advances, whereas systematic cross-architecture comparisons, controlled IID/OOD assessments, and explicit validation against device-relevant equilibrium references remain comparatively limited.

To address these gaps, we present a systematic and controlled comparison of five representative neural architectures—MLP, CNN, FNO, Transformer, and KAN—for approximating the mapping from input plasma parameters and geometric descriptors to GS equilibrium solutions. The main contributions are as follows:

Table 1. Comparison of representative learning-based Grad-Shafranov solvers.

| Study | Training Data Sources | Architecture Comparison | Real-Device Validation | Parameter Extrapolation | Sim-to-Real |
|---|---|---|---|---|---|
| **S. Joung et.al** | Real magnetic measurements | ✗ | ✓ | ✗ | ✗ |
| **L. L. Lao et.al** | Experimental reconstruction | ✗ | ✓ | ✗ | ✗ |
| **B. Jang et.al** | Data-free (PDE residual) | ✗ | ✗ | ✓ | ✗ |
| **F. N. Rizqan et.al** | Data-free (PDE residual) | ✓ | ✗ | ✓ | ✗ |
| **C. Zhou et.al** | Data-free (PDE residual) | ✗ | ✗ | ✗ | ✗ |
| **M. Bonotto et.al** | Simulator-generated | ✗ | ✗ | ✗ | ✗ |
| **Z. Wang et.al** | Simulator-generated | ✗ | ✗ | ✗ | ✗ |
| **Ours** | Simulator-generated | ✓ | ✓ | ✓ | ✓ |

To address these gaps, we present a systematic, controlled comparison of representative deep learning architectures spanning the major network paradigms—fully connected networks (MLP), convolutional networks (CNN), spectral operator networks (FNO), attention-based models (Transformer), and learnable-activation networks (KAN) for solving the Grad-Shafranov equation. The main contributions are as follows:

1. A systematic AI surrogate framework with comprehensive architecture benchmarking is established for real-time tokamak equilibrium prediction. 2. A simulation-to-device validation pathway is demonstrated for AI-based equilibrium prediction. 3. Scaling behavior and generalization characteristics of AI surrogate models are investigated.

This paper is organized as follows. Section 2 introduces the problem formulation and the construction of the high-fidelity equilibrium dataset. Section 3 describes the surrogate architectures together with the corresponding training strategies and evaluation criteria. Section 4 presents the experimental results, including prediction accuracy comparison, generalization analysis, scaling behavior, and out-of-distribution extrapolation performance. Finally, Section 5 summarizes the main findings and discusses future research directions.

## 2. Tokamak Equilibrium Prediction Problem

## 2.1 Background and Motivation

In the standard cylindrical coordinate system $(R, \phi, Z)$, the axisymmetry of a tokamak plasma implies that all physical quantities are independent of the toroidal angle $\phi$. Under this assumption, the equilibrium configuration can be described by the poloidal flux function $\psi(R, Z)$, whose contours define the magnetic flux surfaces. The Grad-Shafranov (GS) equation, derived from the ideal magnetohydrodynamic (MHD) equilibrium condition $\nabla P = J \times B$, can be written in its standard form as:

$$\Delta^* \psi = -\mu_0 R^2 \frac{dp}{d\psi} - F \frac{dF}{d\psi} \tag{1}$$

Where p(ψ) is the plasma pressure, and $F(\psi)=RB_T$ is a function related to the poloidal current, where $B_T$ is the toroidal magnetic field. Both the pressure p(ψ) and the function F(ψ) are free functions that must be prescribed for a given equilibrium. The Δ* operator is defined as:

$$\Delta^* \psi = \frac{\partial^2 \psi}{\partial R^2} - \frac{1}{R}\frac{\partial \psi}{\partial R} + \frac{\partial^2 \psi}{\partial Z^2} = R \frac{\partial}{\partial R}\left(\frac{1}{R}\frac{\partial \psi}{\partial R}\right) + \frac{\partial^2 \psi}{\partial Z^2}$$

We use the normalized flux $\hat{\psi}$ and prescribe the pressure and poloidal-current source terms with the profile families defined in Eqs. (3) and (4). The pressure-gradient contribution and the poloidal-current contribution jointly determine the toroidal current-density distribution and hence the internal flux-surface structure. Normalization fixes $\hat{\psi}=0$ at the magnetic axis and $\hat{\psi}=1$ at the plasma boundary, allowing equilibria with different currents and dimensions to be represented on a common numerical scale. Here $I_p$ is the total plasma current; $n_p$, $m_p$, $n_f$, and mf control the pressure and current-profile shapes. Their ranges are summarized in Table 2.

$$\hat{\psi} = \frac{\psi - \psi_0}{\psi_a - \psi_0} \tag{2}$$

$$p'(\psi) = \beta_0 R_0 (1 - \hat{\psi}^{n_p})^{m_p} \tag{3}$$

$$FF'(\psi) = (1 - \beta_0) R_0 \lambda \hat{\psi}^{n_f} (1 - \hat{\psi}^{n_f})^{m_f} \tag{4}$$

where $R_0$ denotes the major radius of the device; the current profile is characterized by the free parameters $n_p$, $m_p$, $n_f$, $m_f$, $\lambda$ and $\beta_0$; the indices $n_p$, $m_p$, $n_f$ and $m_f$ control the shape of the pressure and poloidal current profiles, while $\lambda$ is associated with the total plasma current $I_p$; the parameter $\beta_0$ typically takes a value of 0.8.

The fixed plasma boundary is parameterized by major radius $R_0$, minor radius $a$, elongation $\kappa$, and triangularity $\delta$:

$$R(\tau) = R_0 + a\cos(\tau + \delta \sin \tau) \tag{5}$$

$$Z(\tau) = \kappa a \sin \tau \tag{6}$$

For fixed-boundary equilibrium problems, the boundary condition is imposed by setting the poloidal flux on the plasma boundary $\Gamma$ to zero:

$$\psi(R, Z)|_{(R,Z)\in\Gamma} = 0 \tag{7}$$

## 2.2 Numerical Solution Method

To generate high-fidelity reference solutions for training and evaluating the AI surrogate models, we numerically solve the Grad–Shafranov (GS) equation using a finite-difference method (FDM) combined with successive over-relaxation (SOR) [31]. The computational domain is the (*R*, *Z*) poloidal plane, discretized using a uniform rectangular grid of 101×161 point. Iteration stops at the same convergence tolerance used for the original dataset. The current MATLAB CPU baseline requires 10.2 s per case; no unrecorded relaxation factor or iteration cap is introduced here. At every iteration, the interior flux is updated from its four nearest grid neighbors and the nonlinear source term, while boundary nodes remain fixed at $\psi$=0. This baseline is used both to generate supervised targets and to define the latency comparison; it is not presented as the fastest available GS solver.

## 2.3 Dataset Preparation

To enable robust training and rigorous evaluation of the surrogate models, three datasets were generated using a MATLAB-based Grad-Shafranov equilibrium solver. A central consideration in model assessment is the distinction between Independent and Identically Distributed (IID) generalization and Out-of-Distribution (OOD) extrapolation. The former evaluates model performance on unseen samples drawn from the same parameter distribution as the training set, whereas the latter assesses robustness on samples whose parameter ranges intentionally deviate from the training domain.

The input parameters are sampled using a combination of Latin hypercube sampling[32] and integer sampling to systematically explore the geometric and physical quantities relevant to plasma equilibrium. The shaping parameters include the major radius $R_0$, minor radius $a$, elongation $\kappa$, and triangularity $\delta$; the plasma current $I_p$ and four free coefficients $n_p$, $m_p$, $n_f$, $m_f$ that determine the pressure and current profile shapes. This design ensures uniform coverage and efficient exploration of the high-dimensional parameter space associated with the equilibrium physics. The datasets are constructed as follows:

(1) **IID dataset.** This dataset contains 100,000 samples and is used for model training, validation, and in-distribution testing. Among them, 60,000 samples are used for training the surrogate models, 20,000 samples are used for validation during hyperparameter tuning, and 20,000 samples serve as the IID test set for evaluating generalization performance.

(2) **OOD dataset**. To assess the extrapolation capability of the models under shifted physical conditions, an additional dataset of 10,000 samples is constructed for OOD testing. The sampling ranges of selected parameters extend beyond the IID boundaries, forming new regions in the parameter space. This design enables a systematic evaluation of the model adaptability and robustness outside the training distribution.

(3) **EXL-50U experimental validation dataset**. To further validate the practical applicability of the developed surrogate models under real device operating conditions, we additionally conduct experimental validation on the EXL-50U tokamak. Equilibrium prediction results from representative discharges are used to benchmark the model predictions against physics-based GS solutions and the operational Shape Editor (SE) reference, thereby establishing a device-level validation chain beyond purely numerical IID/OOD tests.

Each sample in these datasets contains the corresponding 2D poloidal flux field computed by the numerical solver, providing the mapping from the input parameter space to the magnetic equilibrium. The parameter ranges, dataset partitions, and dataset sizes are summarized in Table 2. Both IID and OOD settings are considered: IID is used to evaluate interpolation generalization within the parameter ranges covered by the training data, whereas OOD is used to evaluate extrapolation generalization beyond the training distribution.

Table 2. Parameter ranges, dataset composition, and intended usage

| Dataset | IID | | | OOD |
|---|---|---|---|---|
| | Training set | Validation set | Test set | Test set |
| Size | 60,000 | 20,000 | 20,000 | 10,000 |
| $R_0$ | [1.00, 1.50] | [1.00, 1.50] | [1.00, 1.50] | [0.65, 2.00] |
| $a$ | [0.30, 0.50] | [0.30, 0.50] | [0.30, 0.50] | [0.10, 0.60] |
| $\kappa$ | [1.00, 2.50] | [1.00, 2.50] | [1.00, 2.50] | [0.50, 3.00] |
| $\delta$ | [-0.50, 0.50] | [-0.50, 0.50] | [-0.50, 0.50] | [-0.90, 0.90] |
| $I_p$ | [$1.00 \times 10^6$, $2.00 \times 10^6$] | [$1.00 \times 10^6$, $2.00 \times 10^6$] | [$1.00 \times 10^6$, $2.00 \times 10^6$] | [$5.00 \times 10^5$, $3.00 \times 10^6$] |
| $n_p$, $m_p$, $n_f$, $m_f$ | [1.00, 8.00] | [1.00, 8.00] | [1.00, 8.00] | [1.00, 16.00] |

## 3. AI Surrogate Modeling Framework

### 3.1 Problem Formulation and Evaluation Framework

The objective of the present study is to construct fast and accurate AI surrogate models for predicting Grad-Shafranov (GS) equilibrium solutions under varying plasma parameters and boundary geometries. Let

$$f = \left(R_0, a, \kappa, \delta, I_p, n_p, m_p, n_f, m_f\right) \tag{8}$$

denote the input parameter vector, and let $\mathcal{G} = \{(R_j, Z_k)\}$ denote the computational grid in the poloidal plane. For a given $\xi$, the high-fidelity numerical GS solver produces the corresponding two-dimensional poloidal-flux field $u$. The surrogate-prediction task is therefore formulated as

$$\hat{u}_\theta = \mathcal{S}_\theta(f, \mathcal{G}) \tag{9}$$

where $\mathcal{S}_\theta$ denotes an AI surrogate model with trainable parameters $\theta$, and $\hat{u}_\theta$ is the predicted poloidal-flux field. Rather than assuming that all candidate models are neural operators, this formulation provides a common parameter-to-field prediction task under which different AI surrogate architectures can be compared fairly.

**(1) High-fidelity dataset generation**

In the offline stage, input parameters are sampled from the prescribed geometric and equilibrium-profile parameter ranges. For each parameter vector $f_i$, the numerical GS solver computes the corresponding reference poloidal-flux field $u_i$. The resulting dataset is written as:

$$\mathrm{D} = \{(f_i, u_i)\}_{i=1}^{N} \tag{10}$$

The numerical solutions provide common reference targets for all five surrogate architectures, ensuring that differences in performance arise primarily from model architecture and training behavior rather than from differences in data generation.

**(2) AI surrogate training**

Each candidate architecture is trained to approximate the mapping from the input parameters and spatial coordinates to the corresponding poloidal-flux field. Although the architectures process these inputs differently, they use the same training, validation, and testing partitions and are optimized under a unified protocol. The core component of the total loss function $\mathrm{L}(\theta)$ is the data-misfit term $\mathrm{L}_{data}$. For a given training dataset $\{f_i, u_i\}_{i=1}^{N}$, it's defined as:

$$\mathrm{L}_{\mathrm{data}} = \frac{1}{N}\sum_{i=1}^{N} |\psi_{\mathrm{pred}} - \psi_{\mathrm{true}}|^2 \tag{11}$$

**(3) Model validation**

The trained models are evaluated from four complementary perspectives: predictive accuracy, computational efficiency, scaling behavior, and robustness. Generalization is assessed using both IID test samples drawn from the same parameter distribution as the training data and OOD samples containing previously unseen parameter ranges and plasma geometries. Relative error metrics are used to quantify field-prediction accuracy, while parameter count and inference latency are measured to evaluate suitability for real-time deployment. In addition, device-relevant validation is conducted using EXL-50U equilibrium configurations by comparing surrogate predictions with numerical GS solutions and Shape Editor reference equilibria.

This framework provides a unified basis for determining how architectural differences affect accuracy, OOD robustness, and inference efficiency. These considerations are particularly important for real-time tokamak applications, where a surrogate model must provide accurate equilibrium predictions within millisecond or sub-millisecond latency constraints. A schematic illustration of the overall workflow is shown in Fig. 1.

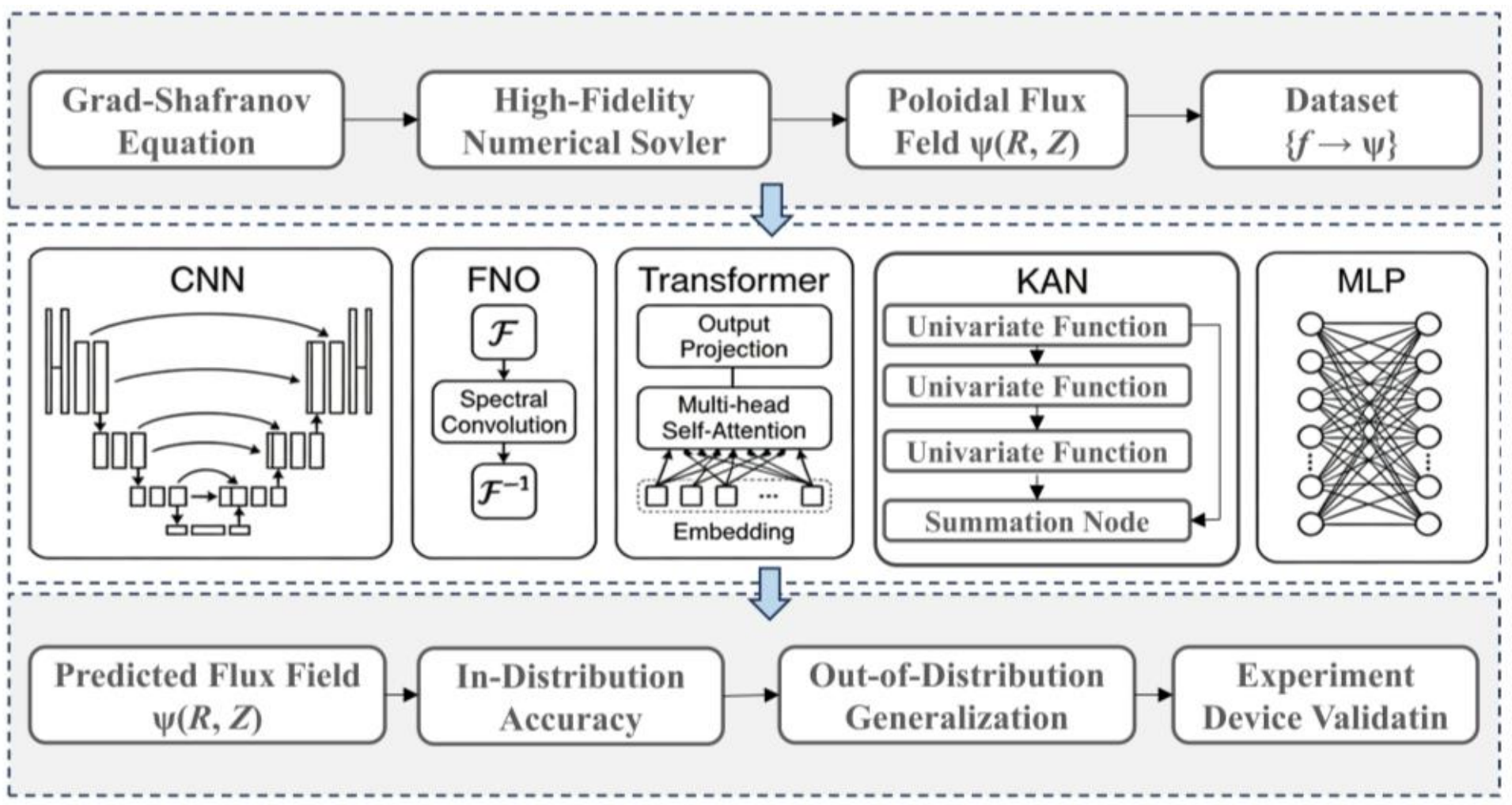

Fig. 1. Overall architecture diagram.

## 3.2 AI surrogate architectures

Five representative AI surrogate architectures (MLP, CNN, FNO, Transformer, and KAN), these are evaluated under the unified framework described above. All models receive the same physical and geometric information and predict the same two-dimensional poloidal-flux field. However, they differ substantially in how they represent spatial coordinates, parameter dependence, local structures, and global interactions. These differences directly affect their prediction accuracy, OOD robustness, computational cost, and suitability for real-time plasma control.

For the GS equation, the inputs consist of the plasma shaping parameters $f = (R_0, a, \kappa, \delta, Ip, np, mp, nf, mf)$ and spatial coordinates (*R*, *Z*), while the output *u* corresponds to the magnetic flux field $\psi$. Although conventional regression networks can perform direct feed-forward prediction, they struggle to represent the complex dependencies inherent in physical fields. To enhance the expressiveness of the surrogate model for the flux distribution, this study conducts a systematic comparison of several neural network architectures. The following sections present the backbone models employed in this work and describe their corresponding training paradigms.

### 3.2.1 Convolutional Neural Network

The CNN surrogate adopts a U-Net-style encoder–decoder architecture. Its input is a multichannel grid containing the spatial coordinates and broadcast parameter fields, while its output is the predicted poloidal-flux field on the same grid. The encoder applies convolution and downsampling operations to extract multiscale spatial features. The decoder then restores the original resolution through upsampling operations, while skip connections transfer fine-scale information from the encoder to the corresponding decoder layers.

This architecture is well suited to the spatially structured nature of the GS solution. Local convolutional kernels capture flux gradients and boundary-related features, whereas the multiscale encoder–decoder structure expands the effective receptive field and facilitates reconstruction of the global equilibrium configuration. Convolutional operations are also highly optimized on modern GPUs and inference engines, making the CNN a strong candidate for real-time deployment. Its principal limitation is that its spatial representation is tied to the discretized grid used during training.

Convolutional Neural Networks (CNNs) are a class of deep learning models designed for structured grid data. Their fundamental principle lies in extracting local spatial features through convolutional kernels and building hierarchical feature representations via stacked convolution and pooling layers. In this study, the spatial coordinates ($R$, $Z$) are discretized onto a regular mesh and reformulated as a two-channel pseudo-image, where each pixel stores the corresponding $R$ and $Z$ coordinate values. This structured representation enables the model to learn the mapping from the coordinate domain to the magnetic flux distribution in an image-based manner.

A U-Net architecture is adopted as the CNN backbone (see Fig. 2). The network follows a symmetric encoder-decoder design, incorporating skip connections between layers of equal resolution to fuse high-level semantic features with low-level spatial information. The encoder progressively extracts multiscale spatial features through convolution and pooling operations, while the decoder progressively restores spatial resolution via up-sampling and convolution, yielding a high-fidelity prediction of the magnetic flux field:

$$Z_{m+1} = \sigma(W_m * Z_m + b_m) \tag{12}$$

$$P_{m+1} = \mathrm{MaxPool}(Z_{m+1}, h) \tag{13}$$

where $Z_m$ and $Z_{m+1}$ denote the input feature map of layer $m$ and the output feature map of layer $m$+1, respectively; $*$ represents the convolution operation; $W_m$ and $b_m$ are the learnable weights and biases; $h$ and $\sigma$ denote the pooling window size the activation function.

During decoding, the spatial resolution is progressively recovered through up-sampling operations (e.g., transposed convolution).

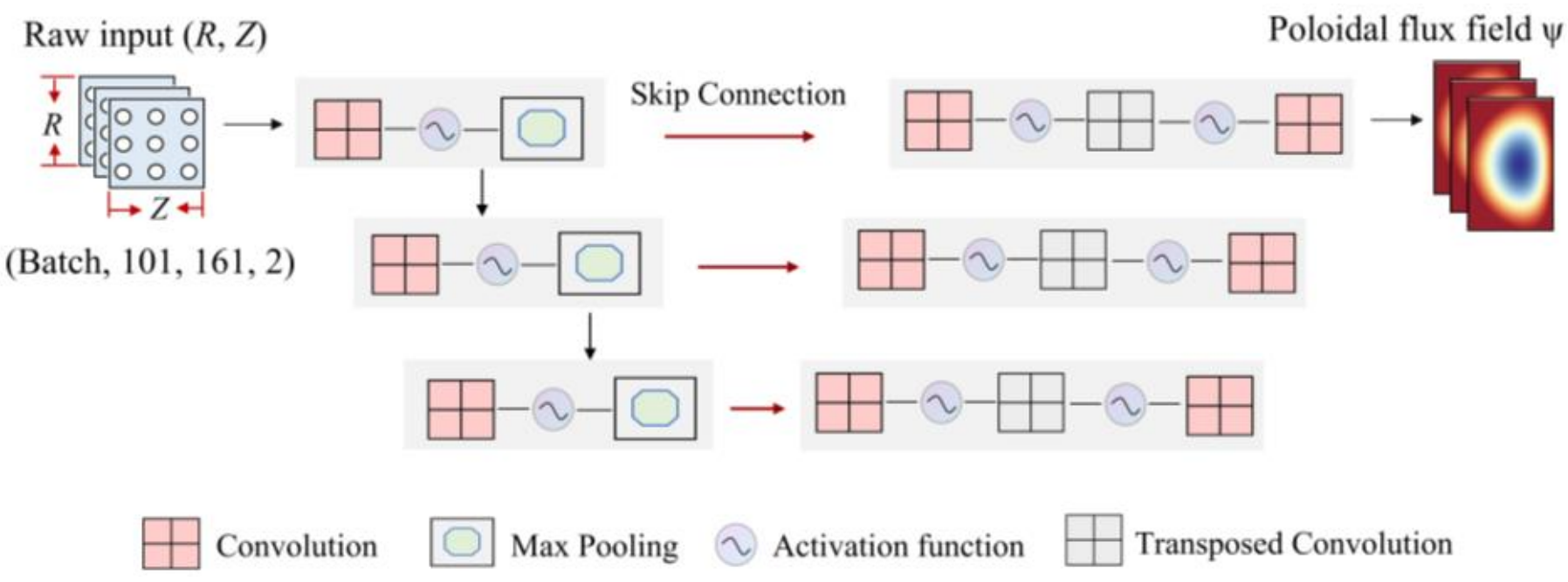

Fig. 2. Architecture of the CNN.

### 3.2.2 Fourier Neural Operator

The Fourier Neural Operator (FNO) [23] is designed to learn mappings between infinite-dimensional function spaces and represents a prominent framework for operator learning. Unlike conventional neural networks that rely on fixed discretization grids, FNO exhibits discretization invariance, enabling inference across different mesh resolutions without retraining. This property provides strong cross-grid generalization, and the overall architecture is illustrated in Fig. 3.

The core mechanism of FNO is its use of global convolution in Fourier space. According to the convolution theorem, a convolution operation in the spatial domain is equivalent to pointwise multiplication in the spectral domain, as expressed in Eq. (14):

$$(f * g)(x) = \mathcal{F}^{-1}\left[\mathcal{F}[f] \cdot \mathcal{F}[g]\right](x) \tag{14}$$

where $\mathcal{F}$ and $\mathcal{F}^{-1}$ denote the Fourier transform and inverse Fourier transform, respectively. Building on this principle, each FNO layer performs linear feature transformation and mode truncation in frequency space, followed by an inverse Fourier transform back to the spatial domain. The layer-wise update rule can be written as:

$$v_{t+1}(x) = \sigma\left(\mathcal{F}^{-1}\left[R_\phi \cdot \mathcal{F}[v_t]\right](x) + W v_t(x)\right) \tag{15}$$

where $v_{t+1}(x)$ is the feature representation at layer $t$; $\mathcal{F}\left[v_t\right]$ denotes its spectral coefficients in the Fourier domain; $R_\phi$ is the learnable linear transformation matrix in frequency space and $W$ is a trainable linear mapping applied in the spatial domain.

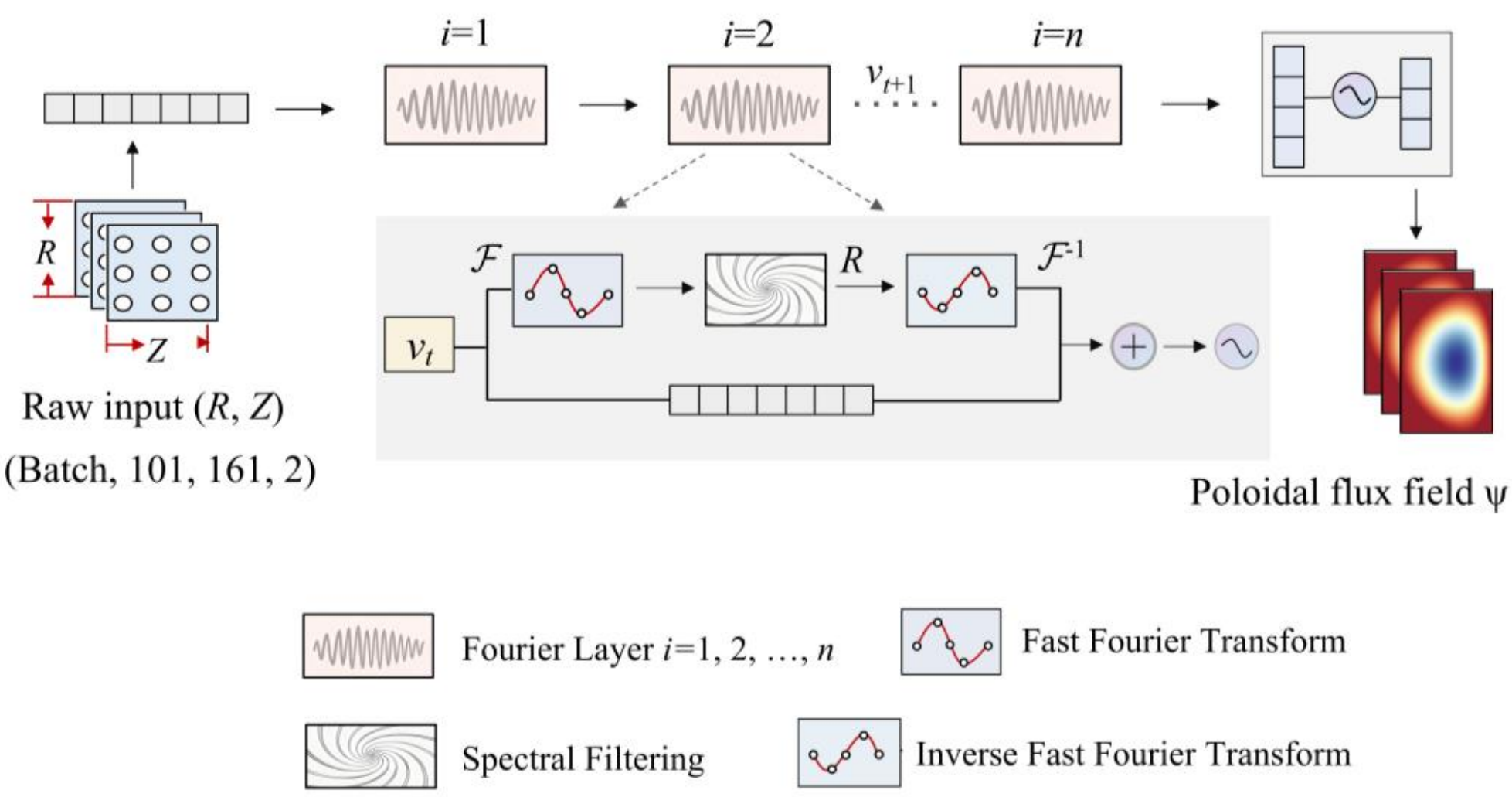

Fig. 3. Architecture of the FNO.

### 3.2.3 Kolmogorov-Arnold

The Kolmogorov-Arnold Network (KAN), inspired by the Kolmogorov-Arnold representation theorem, replaces the fixed activation functions in conventional MLPs with learnable univariate functions, as shown in Fig. 4. For an input vector $x = (x_1, \ldots, x_n)$, the $j$-th output can be written as:

$$y_j = \sum_{i=1}^{n} \phi_{j,i}(x_i) \tag{16}$$

where $\phi_{j,i}$ denotes the learnable univariate functions, which are typically constructed using B-spline basis functions together with trainable coefficients. This formulation enables explicit learning of input-output mappings in a dimension-wise manner, allowing the model to maintain strong expressive capacity while offering improved interpretability. In addition, KAN achieves high-dimensional function approximation through hierarchical compositions of such univariate mappings, thereby providing a balance between approximation accuracy and model transparency.

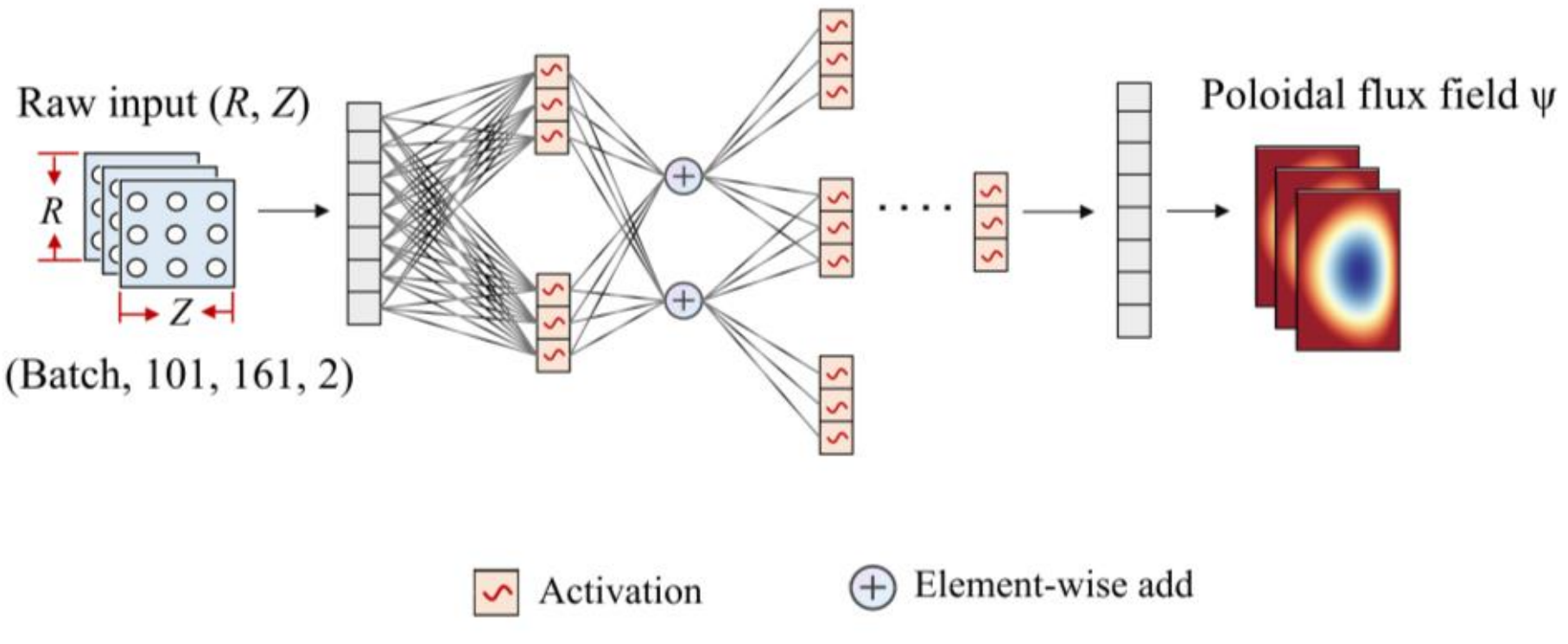

Fig. 4. Architecture of the KAN.

### 3.2.4 Transformer

The Transformer models [24] global dependencies through the self-attention mechanism, enabling it to effectively capture long-range interactions across different positions in the input sequence. Therefore, the model has been increasingly adopted in physics-related problems and scientific computing in recent years to represent physical processes with global coupling characteristics [24]. In this work, the GS equation is formulated as a sequence-to-field mapping problem, where the geometric shaping parameters and spatial coordinates are jointly encoded as input tokens. After embedding and positional encoding, these tokens are fed into Transformer blocks, as illustrated in Fig. 5.

Within the encoder, the multi-head self-attention mechanism learns global physical correlations through weighted interactions among features. The decoder incorporates target-position embeddings to reconstruct the solution. Finally, the spatial representations are mapped to the desired physical field through a linear layer or Fourier layer, thereby achieving efficient prediction of the magnetic flux function from geometric parameters and spatial locations.

The scaled dot-product attention is computed as follows:

$$\text{Attention}(Q,K,V)m,n = \frac{\sum i = 1^{S} \phi_q(Q_{m,i})\phi_k(K_{n,i})V_{n,i}}{\sum_{i=1}^{S} \phi_q(Q_{m,i})\phi_k(K_{n,i})} \tag{17}$$

where **Q**, **K**, **V** denote the Query, Key, and Value matrices, respectively; $d_k$ is the feature dimension of each attention head; $S$, $m$, $n$ are the sequence length, index the Query and Key positions, respectively; $\phi_k(g)$ is the feature projection function and $V_{n,i}$ represents the value feature at position $i$.

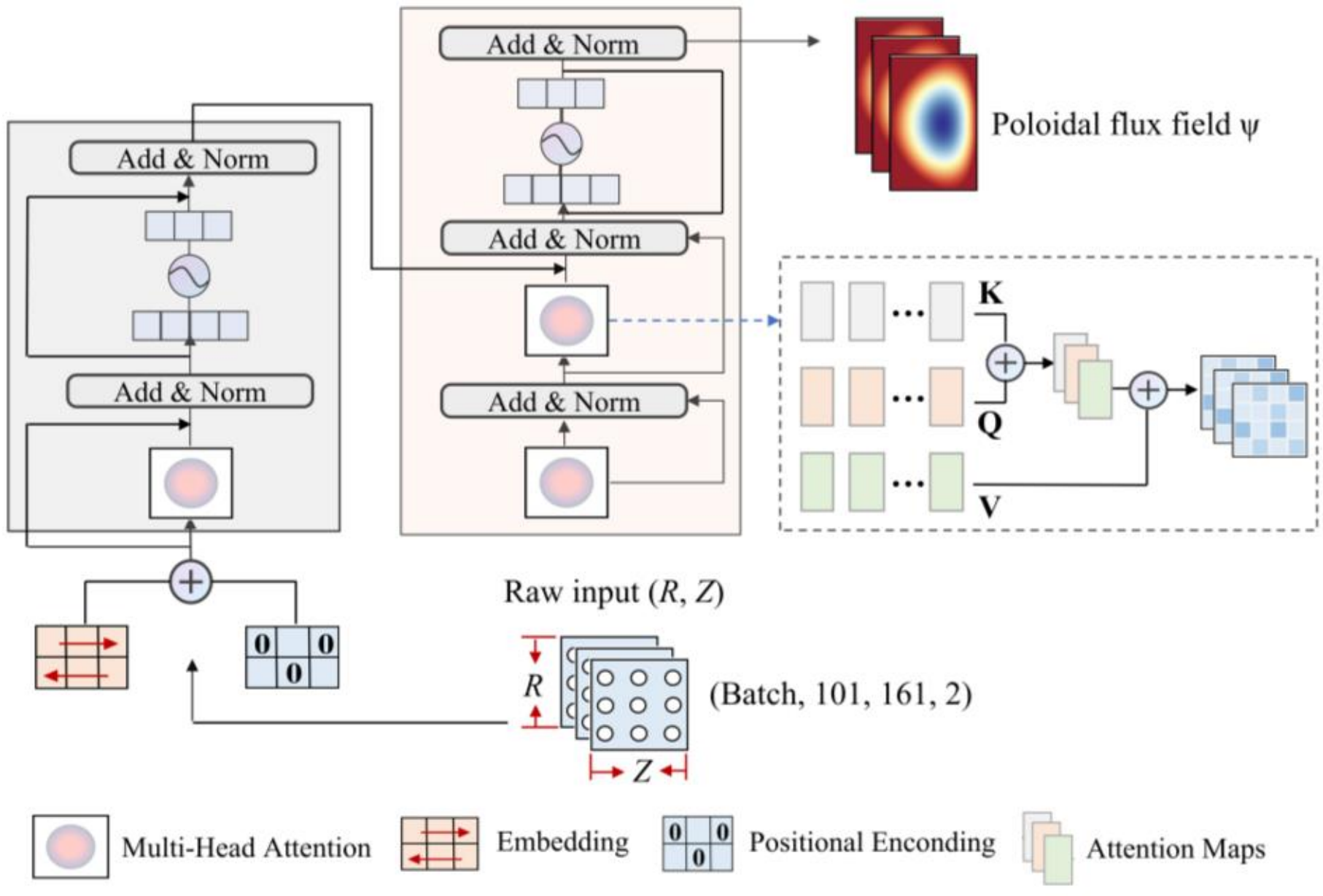

Fig. 5. Architecture of the Transformer.

The five architectures represent distinct strategies for GS equilibrium prediction. The MLP provides a simple pointwise baseline; the CNN captures multiscale spatial structures through local convolution; the FNO represents global equilibrium behavior through spectral convolution; the Transformer models long-range spatial interactions using self-attention; and the KAN represents nonlinear parameter dependence through learnable univariate functions. Their comparison under identical datasets and evaluation protocols enables a direct assessment of the trade-offs among accuracy, inference efficiency, model capacity, and OOD robustness, thereby providing practical guidance for selecting AI surrogate architectures for real-time tokamak equilibrium prediction.

Table 3. Summary of model architectures, input/output processing, parameter scale, and key characteristics

| Model type | Input processing | Backbone architecture | Output architecture | Parameters | Key features |
|---|---|---|---|---|---|
| MLP | Coordinate-parameter concatenation | 6 fully connected layers | Direct field output | ~$10^5$ | Baseline model; simple and lightweight |
| CNN | 2D grid representation | Encoder-Decoder | 2D field prediction | $\sim 10^5$ | Rich hierarchical features; skip connections facilitate localization |
| FNO | Fourier transform of inputs | Spectral convolution layers | Inverse transform | $\sim 10^5$ | Global receptive field; efficient spectral modeling |
| KAN | Spline activation | Learnable functional layers | Linear combination | $\sim 10^4$ | High interpretability; strong function-approximation capability |
| Transformer | Sequence + positional encoding | Self-attention mechanism | Sequence-to-field projection | $\sim 10^5$ | Long-range dependency modeling; parallelizable computation |

### 3.3 Training Strategy and Evaluation Metrics

To systematically evaluate the capability of different neural models in predicting GS equilibrium solutions, we establish a unified training and evaluation protocol. All models are trained on a workstation equipped with seven NVIDIA RTX 4090 GPUs (24 GB each), using PyTorch 2.0 with Distributed Data Parallel. Training is performed on 60,000 samples for 200 epochs using the Adam optimizer ($\beta_1$=0.99, $\beta_2$=0.99, $\varepsilon$=$10^{-8}$). The initial learning rate $\alpha_0$ is set to $10^{-3}$ for MLP, FNO, Transformer, KAN and $10^{-4}$ for CNNs, with a StepLR decay of 0.1 every 150 epochs. The batch size is 64, except for the memory-intensive KAN, which uses a batch size of 16.

During training, the mean squared error (MSE) is used as the primary loss function to measure the discrepancy between the predicted poloidal flux $\psi_{\text{pred}}$ and the reference ground truth flux $\psi_{\text{true}}$. It reflects the model's accuracy in reproducing the overall flux distribution and is defined as:

$$L_{\text{MSE}} = \frac{1}{N}\sum \left\| \psi_{\text{pred}} - \psi_{\text{true}} \right\|^2 \tag{18}$$

where $N$ denotes the number of samples and $\|g\|$ represents the Euclidean norm for measuring vector or field-level discrepancies.

After training, the numerical accuracy of the predicted flux fields is further assessed. Specifically, we quantify the consistency between the predicted two-dimensional flux $\psi_{\text{pred}}$ and the high-fidelity numerical solution $\psi_{\text{true}}$. Two commonly used relative error metrics are adopted: the mean relative error $L_1(\psi)$ and the root-mean-square relative error $L_2(\psi)$:

$$L_1(\psi) = \frac{\sum_{j=1}^{N} |\psi_{\text{pred}}(\mathbf{x}_j) - \psi_{\text{true}}(\mathbf{x}_j)|}{\sum_{j=1}^{N} |\psi_{\text{true}}(\mathbf{x}_j)|} \tag{19}$$

$$L_2(\psi) = \frac{\sqrt{\sum_{j=1}^{N} |\psi_{\text{pred}}(\mathbf{x}_j) - \psi_{\text{true}}(\mathbf{x}_j)|^2}}{\sqrt{\sum_{j=1}^{N} |\psi_{\text{true}}(\mathbf{x}_j)|^2}} \tag{20}$$

The $L_1(\psi)$ quantifies the overall amplitude deviation between the predicted and reference solutions, whereas the $L_2(\psi)$ emphasizes larger local discrepancies. Owing to the squared term, RMSRE is more sensitive to regions with strong flux gradients.

# 4. Results and Analysis

## 4.1 Accuracy Evaluation of AI Equilibrium Surrogates

### 4.1.1 Training Convergence and IID Prediction Accuracy

This section evaluates the performance of AI surrogate architectures for predicting the magnetic flux function ψ. All models are trained in a supervised learning setting by minimizing a data-driven loss. Five representative backbone architectures, MLP, KAN, FNO, CNN, and Transformer, are systematically compared to assess their suitability and effectiveness for magnetic equilibrium modeling.

To examine the optimization dynamics and learning efficiency of each model, Fig. 6 presents the convergence curves of the validation loss during training. All architectures exhibit rapid error reduction in the early stage (within the first 50 epochs) and reach convergence after approximately 150 epochs, achieving final error levels on the order of $10^{-7}$. The Transformer, KAN, FNO, and CNN models show smooth and stable convergence, whereas the MLP displays mild oscillations in later epochs but ultimately converges. These results indicate that model accuracy is closely linked to representational capacity, and higher-performing architectures generally require longer training times.

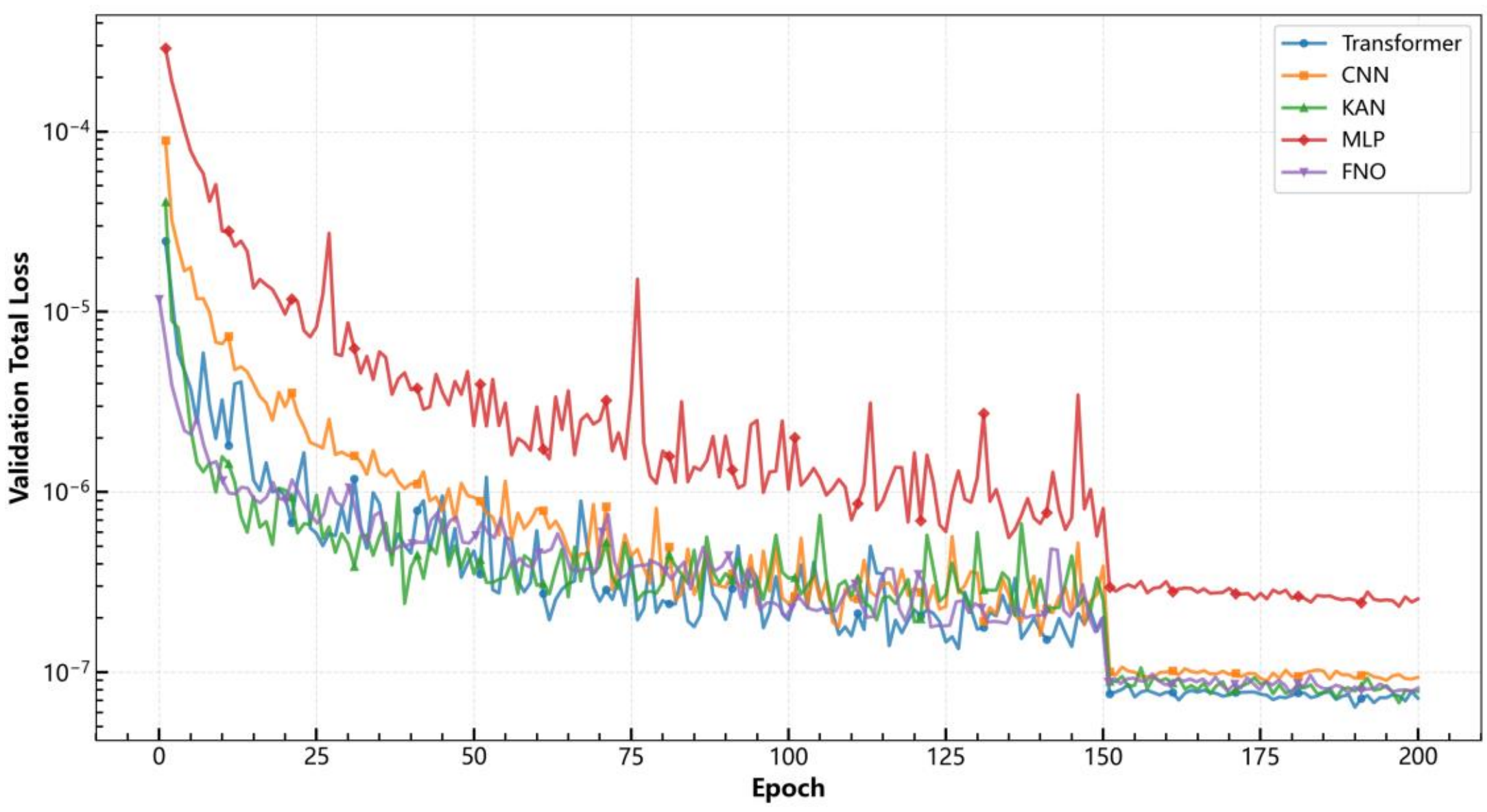

Fig. 6. Training loss convergence of different surrogate models.

After characterizing the convergence behavior of each model, we further perform a quantitative comparison of their predictive performance to comprehensively assess the modeling capability of different network architectures. Table 4 summarizes the detailed results on both the training and test sets. The Transformer achieves the highest accuracy, with test $L_1$ and $L_2$ errors of 0.201% and 0.242%, respectively. The KAN, FNO and CNN models follow closely, yielding comparable performance with test $L_1$ errors of 0.220%, 0.236% and 0.246%. In contrast, the MLP exhibits relatively larger prediction errors. For all models, the training and test errors are nearly identical, with differences below 0.02 percentage points, indicating strong generalization capability and the absence of overfitting. In terms of computational efficiency, substantial differences are observed across architectures. The MLP achieves the shortest training time (5 h 24 min), whereas the KAN (66 h 48 min) and Transformer (50 h 35 min) incur the highest training costs due to their increased model complexity.

Table 4. Performance comparison of different data-driven models

| Model | Training $L_1$ ($\psi$) | Training $L_2$ ($\psi$) | Training Time | Test $L_1$ ($\psi$) | Test $L_2$ ($\psi$) |
|---|---|---|---|---|---|
| MLP | 0.408% | 0.442% | 5h 24m | 0.411% | 0.444% |
| KAN | 0.218% | 0.252% | 66h 48m | 0.220% | 0.253% |

| FNO | 0.233% | 0.262% | 12h 8m | 0.236% | 0.266% |
|---|---|---|---|---|---|
| CNN | 0.242% | 0.281% | 12h 42m | 0.246% | 0.284% |
| Transformer | 0.188% | 0.230% | 50h 35m | 0.201% | 0.242% |

When jointly considering accuracy and computational efficiency, the Transformer achieves the highest IID accuracy. The KAN, FNO, and CNN achieve comparable accuracy whereas the MLP exhibits relatively larger errors. For all models, the differences between the training and test errors remain below 0.02 percentage points, indicating limited generalization gaps under IID test conditions.

### 4.1.2 Statistical Analysis of Prediction Errors on IID

To further assess the predictive performance of the models, Fig. 7 presents the histograms and cumulative distribution functions of the prediction errors for all five architectures. For an ideal predictor, the error distribution should exhibit a sharp and narrow peak centered at zero, indicating high accuracy and the absence of systematic bias. As shown, the mean error of each model is close to zero, confirming that none of the models exhibits noticeable bias. To quantify the dispersion of these errors, we use the mean absolute error (MAE), defined as the average magnitude of the prediction error.

In contrast, the Transformer, KAN, FNO and CNN models yield more concentrated error distributions, each achieving an MAE of 0.0002. Among them, the Transformer exhibits the sharpest peak, indicating that the vast majority of its prediction errors lie within an extremely narrow neighborhood around zero, leading to the highest overall accuracy. The KAN, FNO and CNN show similarly concentrated distributions, whereas the MLP performs intermediately between these models (MAE = 0.0003).

Overall, the visualization of the error distributions aligns closely with the quantitative metrics in Table 4, collectively confirming the superior predictive accuracy of the Transformer, KAN, FNO and CNN architectures for this task.

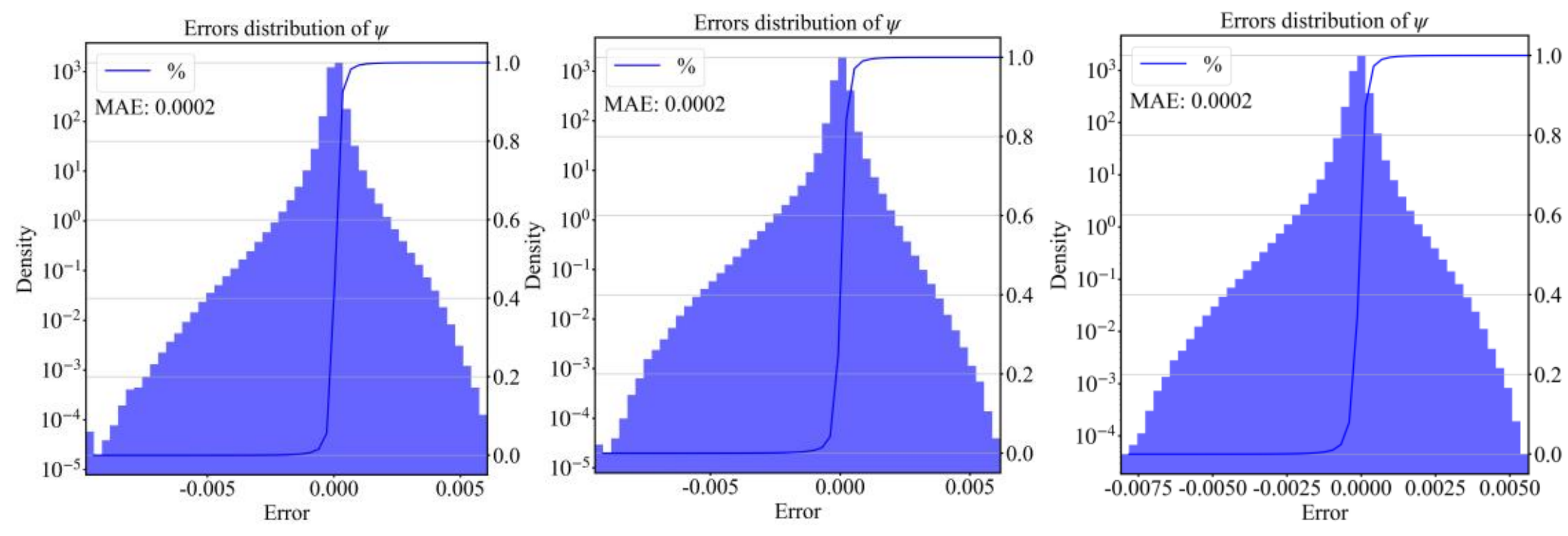

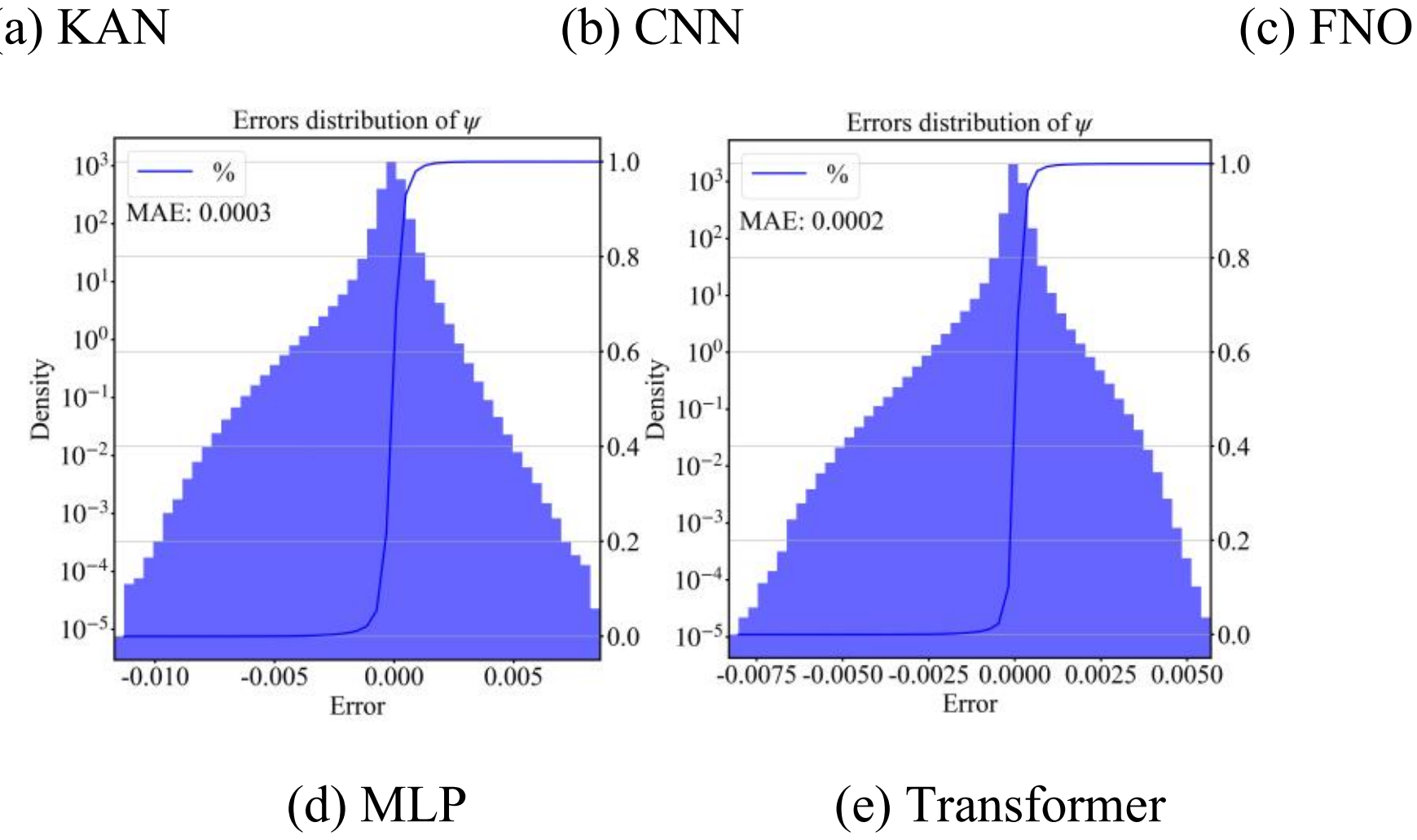

Fig. 7. Error distributions of the predicted magnetic flux ψ for different surrogate models.

### 4.1.3 Visualization of Representative Cases

To complement the quantitative error analysis, a qualitative comparison is conducted between the predicted and reference ground-truth magnetic-flux contours for two representative plasma configurations. As illustrated in Fig. 8 and Fig. 9, the contour distributions corresponding to circular and D-shaped plasmas are visualized for all models. Across all configurations, the predicted magnetic-flux contours closely match the true solutions, accurately capturing the complex topological features associated with different equilibrium shapes. Although the quantitative metrics in Table 4 reveal subtle performance differences among the models, for example, the Transformer exhibits a slight numerical advantage, such distinctions are not visually apparent in the contour comparisons. This qualitative evaluation confirms that all examined deep-learning architectures are capable of predicting plasma-equilibrium fields with high fidelity, even under substantial variations in geometric shaping.

Overall, the error-distribution analysis is consistent with the relative-error results reported in Table 4. Under IID conditions, the Transformer achieves the highest overall accuracy, while the KAN, FNO, and CNN also produce concentrated error distributions with limited mean bias.

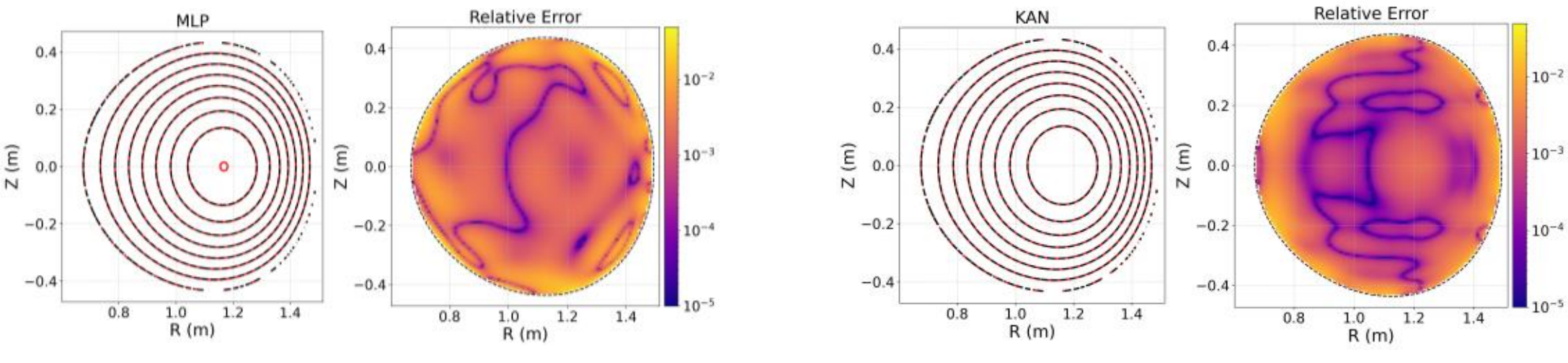

(a) MLP (b) KAN

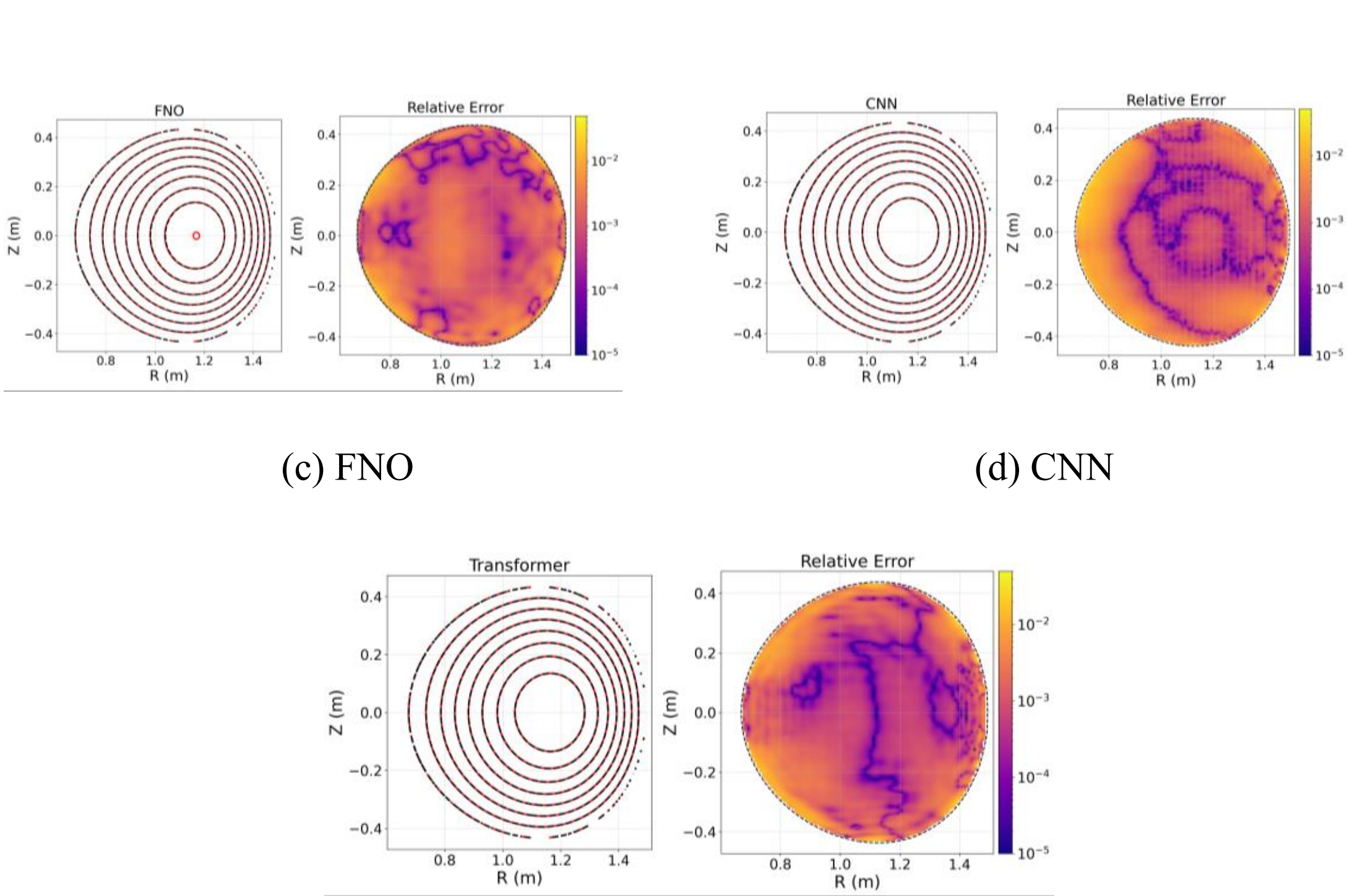

(c) FNO (d) CNN

(e) Transformer

Fig. 8. Circular plasma equilibrium case

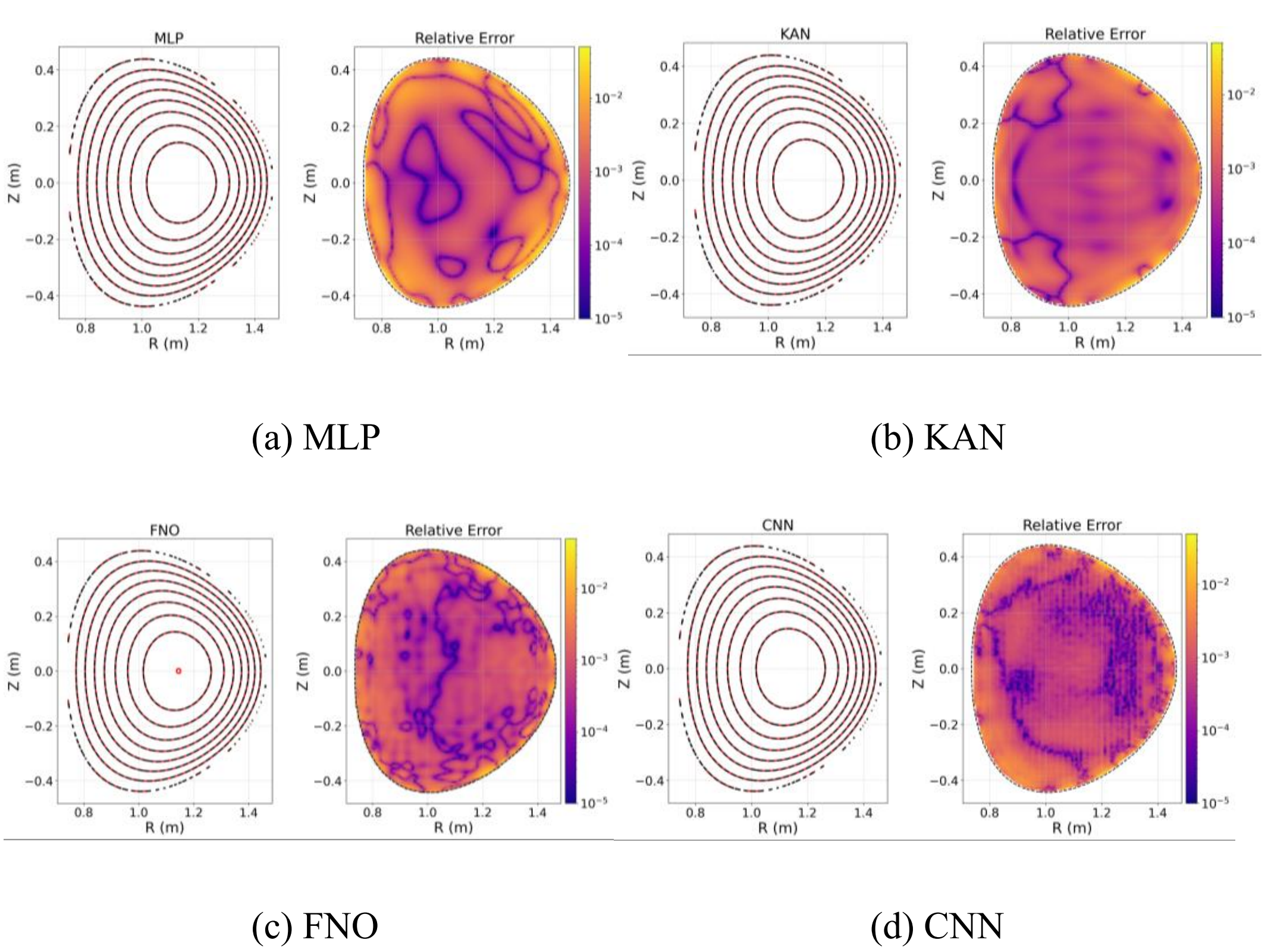

(a) MLP (b) KAN

(c) FNO (d) CNN

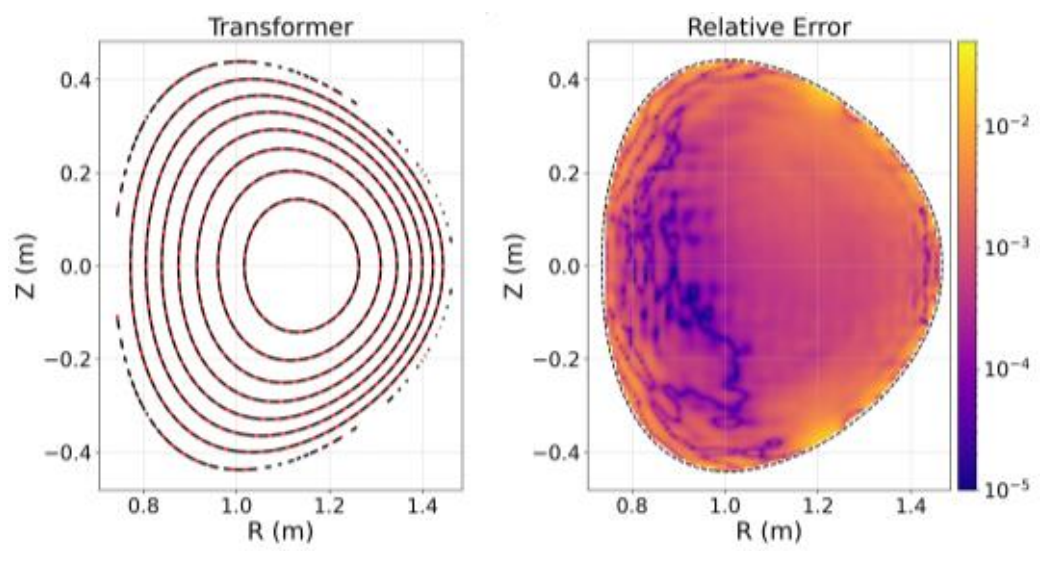

(e) Transformer

Fig. 9. D-Shaped plasma equilibrium case

## 4.2 Robustness Evaluation under IID and OOD Plasma Configurations

### 4.2.1 OOD Generalization Performance

To evaluate the generalization capability of the models beyond the training distribution, stringent out-of-distribution (OOD) test scenarios were constructed. These include: 1) Parameter extrapolation, which examines model performance under plasma parameters outside the ranges observed during training (e.g., pressure-profile and current-profile coefficients); 2) Geometric extrapolation, which assesses the prediction accuracy for plasma shapes not represented in the training data, such as extreme combinations of triangularity and elongation.

The OOD results reveal pronounced differences among the model architectures. As illustrated in Fig. 10 and Fig. 11, despite performance degradation in extrapolation scenarios, the CNN and FNO models still demonstrate the strongest generalization capability, achieving $L_2$ errors of 4.3% and 4.0% respectively, with their predicted magnetic flux contours remaining reasonably consistent with the ground-truth solutions even under extreme geometric configurations not encountered during training. The MLP and Transformer models exhibit moderate extrapolation performance, with $L_2$ errors of 7.8% and 5.8% respectively; notably, the Transformer model experiences significant performance deterioration compared to its superior interpolation accuracy, indicating that although it can preserve the overall magnetic topology structure, its robustness to distributional shifts is relatively weak. In contrast, the KAN exhibit substantial degradation in OOD performance, with $L_2$ errors of 67.2%, and fail to reproduce the plasma boundary and internal flux surfaces accurately. Notably, although the KAN performs well in the in-distribution regime, its performance collapses under extrapolation, suggesting a strong tendency to overfit the training data manifold and a limited capacity to generalize to unseen regions.

(a) MLP (b) KAN

(c) FNO (d) CNN

(e) Transformer

Fig. 10. Circular plasma equilibrium case

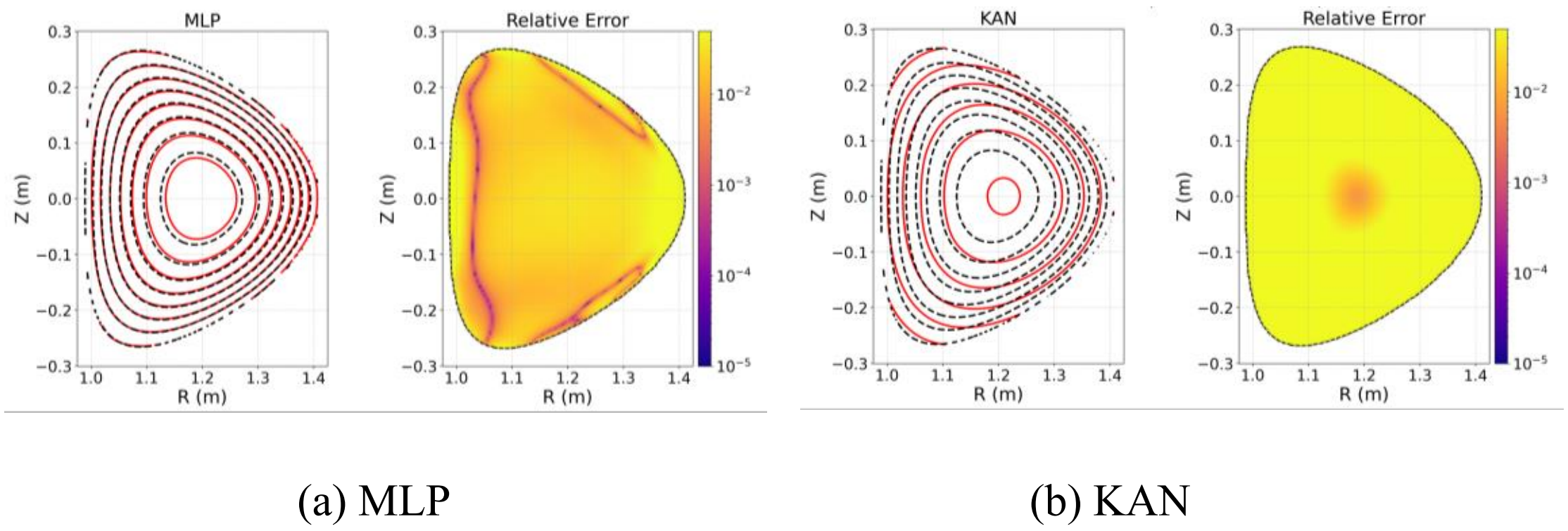

(a) MLP (b) KAN

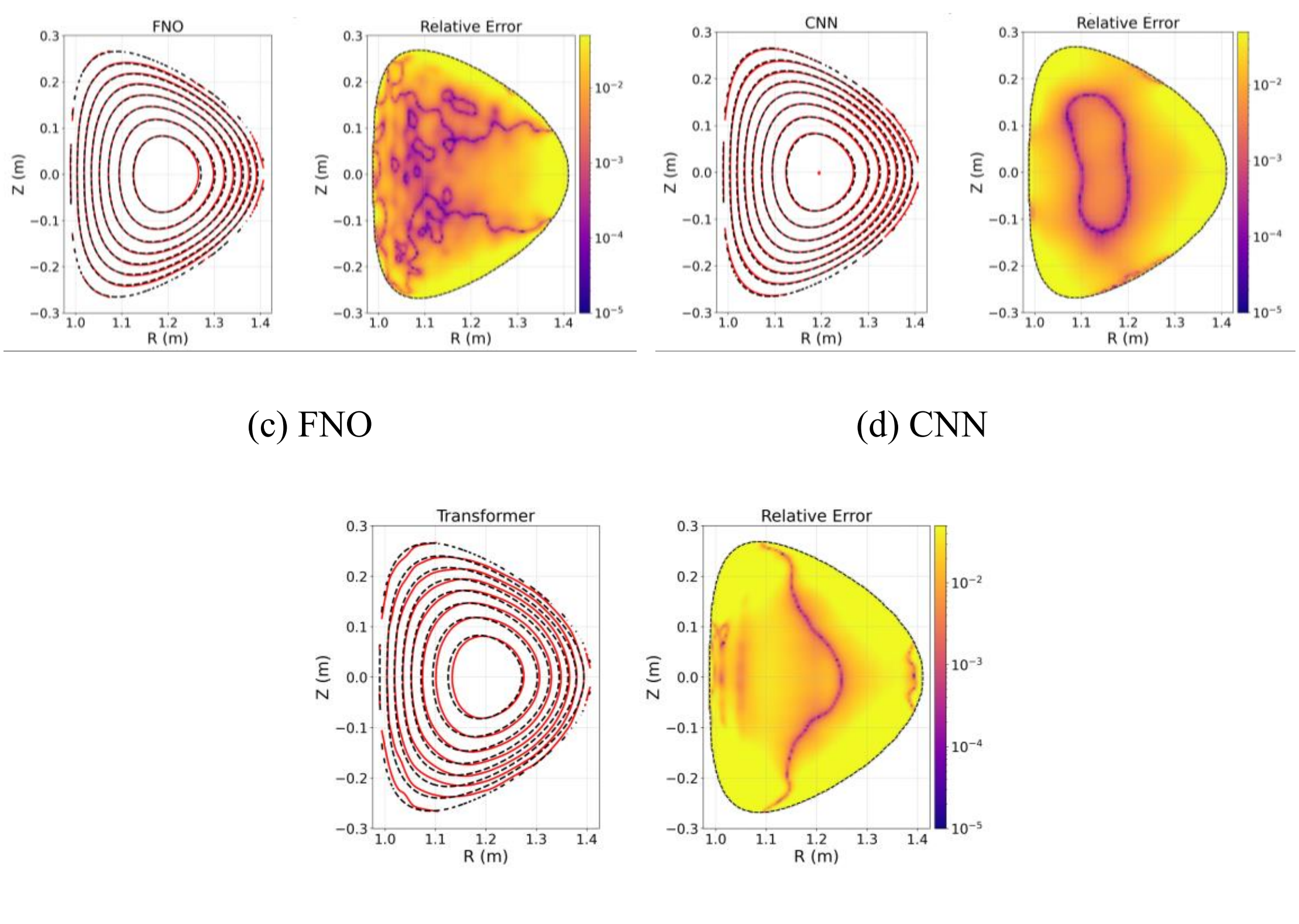

(c) FNO

(d) CNN

(e) Transformer

Fig. 11. D-Shaped plasma equilibrium case

Overall, the OOD experiments clearly indicate that the CNN and FNO architectures provide the most robust extrapolation behavior in this task, benefiting from inherent structural inductive biases that enhance its ability to accommodate distributional shifts in both geometric and physical parameters.

### 4.2.2 Comparison between IID and OOD Performance

A core challenge in developing surrogate models for the Grad-Shafranov equation lies in their generalization capability, specifically their performance on both interpolation and extrapolation of training data. To assess this capability, five neural network architectures CNN, FNO, MLP, KAN, and Transformer are tested on interpolation (in-distribution) and extrapolation (out-of-distribution) tasks. Fig. 12 presents box plot comparisons of $L_1$ and $L_2$ relative errors for each model under both conditions.

All models demonstrate excellent performance on the interpolation task, with $L_1$ relative errors consistently maintained below $5\times10^{-3}$. The Transformer achieves the most accurate predictions, with a median $L_1$ error of $1.95\times10^{-3}$. The small interquartile ranges observed across all models indicate stable and consistent predictions within the training distribution. This uniformly high performance suggests that all architectures possess sufficient capacity to learn the underlying physics-driven mapping when evaluated under input conditions similar to those encountered during training.

The extrapolation task reveals significant disparities in model generalization capabilities, with a clear polarization emerging: CNN, FNO, MLP, and Transformer maintain relatively robust performance (median $L_1$ errors between 0.02 and 0.06), while KAN experiences catastrophic performance degradation (median $L_1$ error approaching 0.7), representing more than two orders of magnitude increase compared to its interpolation performance.

The CNN model demonstrates the strongest extrapolation capability with a median $L_1$ error of 0.027, only 10 times higher than its interpolation error. This resilience can be attributed to the convolutional architecture's inherent inductive biases for spatial locality and translational invariance, which help maintain feature extraction consistency even for out-of-distribution plasma configurations. The FNO achieves comparable extrapolation performance (median $L_1$: 0.031), benefiting from its ability to learn continuous operators in Fourier space that capture global solution characteristics independent of specific discretization or boundary conditions.

The Transformer model, despite its superior interpolation performance, shows only intermediate extrapolation capability (median $L_1$: 0.038). While the self-attention mechanism excels at capturing complex dependencies within the training distribution, it demonstrates reduced robustness to distributional shifts. The increased interquartile range under extrapolation conditions suggests higher prediction uncertainty for novel plasma equilibrium configurations. The MLP, although exhibiting moderate interpolation performance, achieves performance under extrapolation conditions that is slightly inferior to that of the Transformer.

In stark contrast, the KAN model experiences severe generalization failure during extrapolation, with a median $L_1$ error of 0.72. This dramatic performance collapse can be attributed to its architectural limitations. While the KAN theoretically possesses universal approximation capability, its high model flexibility combined with insufficient regularization for physics-constrained problems appears to result in overfitting to the training distribution.

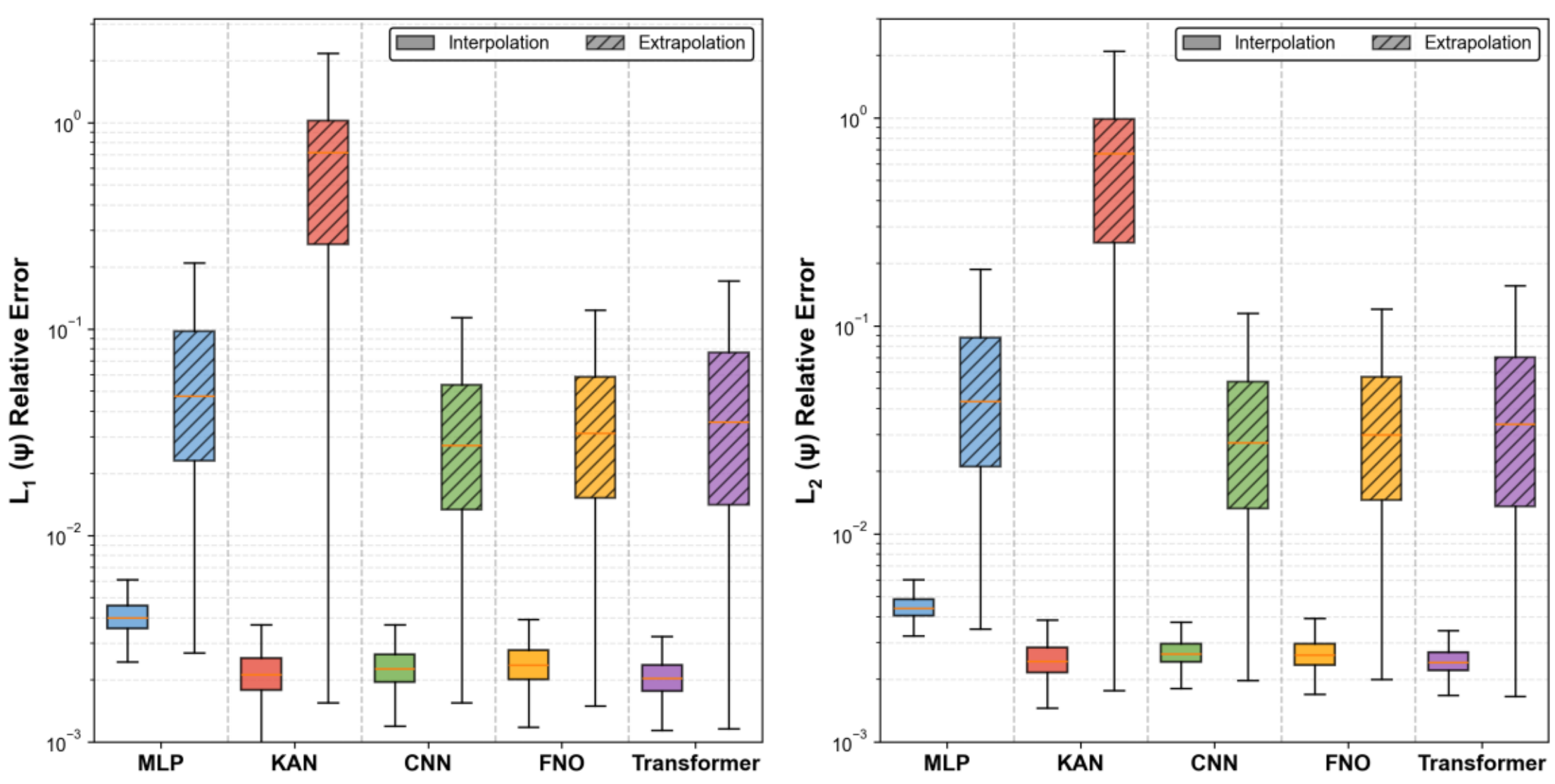

Fig. 12. Model Performance under Interpolation and Extrapolation Conditions.

Overall, the OOD results reveal substantial architecture-dependent differences that are not apparent under IID conditions. The CNN and FNO achieve the strongest OOD performance, whereas the Transformer and MLP exhibit moderate degradation. The KAN, despite its competitive IID accuracy, undergoes severe performance deterioration under extrapolation. These results demonstrate that high IID accuracy does not necessarily imply reliable OOD generalization.

### 4.3 Scaling Behavior of AI Surrogate Models

Given that CNN, FNO, and Transformer demonstrate excellent performance under both interpolation and extrapolation conditions, we further investigated the data efficiency and scalability of these three AI surrogate architectures in predicting Grad-Shafranov equilibrium solutions in solving the Grad-Shafranov equation by evaluating their performance across varying training dataset sizes and model capacities. Specifically, dataset size was varied from 10,000 to 60,000 samples to characterize data utilization efficiency, while network parameter size was scaled from 0.01M to 1M to assess model scalability and parameter efficiency. In contrast, MLP and KAN were excluded from this analysis due to their unsatisfactory performance in either interpolation or extrapolation scenarios. Preliminary evaluations revealed that MLP's limited spatial inductive bias and purely fully-connected architecture fail to adequately capture the complex spatial correlations inherent in magnetic flux distributions. Similarly, KAN, despite its adaptive activation mechanism, exhibited insufficient generalization capability and unstable convergence behavior, resulting in substantially higher prediction errors compared with the three selected architectures. These fundamental limitations rendered MLP and KAN unsuitable for further investigation of data and parameter efficiency. Figures 13(a) and 13(b) present the variation curves of L1 and L2 relative errors as functions of training dataset size under interpolation and extrapolation conditions, respectively, while Figs. 13(c) and 13(d) illustrate the influence of model parameter size under the same evaluation settings.

The interpolation results with varying dataset size (Fig. 13(a)) reveal distinct differences in data utilization characteristics among the three architectures. The Transformer exhibits superior data efficiency and consistently achieves the lowest prediction errors across all dataset sizes. Specifically, for the L1 metric, the Transformer achieves an error of ($4.23\times10^{-3}$) with 10,000 samples and gradually improves to ($2.01\times10^{-3}$) at 60,000 samples, corresponding to a 50% reduction. Similar behavior is observed for L2, demonstrating that the self-attention mechanism effectively exploits limited training data to capture long-range spatial dependencies in magnetic flux distributions. The FNO model shows more pronounced performance gains as the training dataset expands. The most significant improvement occurs during the transition from 10,000 to 40,000 samples, where the L1 error decreases from ($7.23\times10^{-3}$) to ($2.36\times10^{-3}$), representing nearly a 67% reduction. Once sufficient sample coverage is achieved, FNO rapidly approaches the accuracy limit and eventually reaches performance comparable to Transformer. This behavior indicates that stable learning of global spectral representations requires a critical amount of training data. CNN exhibits the strongest dependence on training dataset size. With only 10,000 training samples, CNN produces the highest interpolation error, reaching an L1 error of ($1.04\times10^{-2}$), approximately 2.5 times higher than Transformer. However, increasing the dataset size to 60,000 reduces the error to ($2.46\times10^{-3}$), corresponding to a 76% improvement. This steep learning curve reflects the hierarchical feature extraction mechanism of convolutional networks, which generally requires richer training distributions to construct robust spatial representations. Similar trends observed in both L1 and L2 metrics further confirm the consistency of these findings.

The extrapolation results under data scaling (Fig. 13(b)) reveal substantially different behaviors. FNO demonstrates the strongest extrapolation robustness across all dataset sizes and consistently maintains the lowest prediction errors. With 10,000 training samples, FNO achieves an L1 error of ($6.48\times10^{-2}$), which steadily decreases to ($4.18\times10^{-2}$) at 60,000 samples, corresponding to a 35% improvement. This smooth and monotonic trend suggests that Fourier-domain operator learning provides favorable inductive bias for out-of-distribution generalization. CNN also benefits substantially from larger training datasets in the extrapolation regime. Starting from an L1 error of ($8.73\times10^{-2}$), CNN achieves a 51% reduction and reaches ($4.26\times10^{-2}$) with sufficient training coverage, gradually narrowing the performance gap with FNO. This result suggests that convolutional architectures require broader data distributions to learn transferable local spatial patterns. By contrast, Transformer exhibits weaker extrapolation scalability. Unlike its superior interpolation performance, Transformer shows non-monotonic behavior, where the L1 error fluctuates from ($6\times10^{-2}$) to ($8\times10^{-2}$) before decreasing to ($6.48\times10^{-2}$). This phenomenon suggests that attention-based representations, while highly effective for in-distribution learning, may be more sensitive to distribution shifts and less capable of translating additional training samples into improved extrapolation performance.

Figures 13(c) and 13(d) further investigate model scaling behavior by varying network parameter size. Under interpolation conditions (Fig. 13(c)), all three architectures benefit to varying degrees from increased model capacity. Transformer demonstrates the highest parameter efficiency. Across the entire parameter range from 0.01M to 1M, the L1 error changes only from ($2.006\times10^{-3}$) to ($1.935\times10^{-3}$), corresponding to a modest 6.1% improvement, while the L2 error varies marginally from ($2.414\times10^{-3}$) to ($2.395\times10^{-3}$). These results indicate that Transformer reaches sufficient representational power even at small scales and derives limited additional benefit from increasing parameter count. FNO exhibits the strongest interpolation scalability. As parameter size increases from 0.01M to 1M, the L1 error decreases from ($2.739\times10^{-3}$) to ($1.566\times10^{-3}$), corresponding to a 42.8% reduction, while the L2 error decreases from ($2.966\times10^{-3}$) to ($1.928\times10^{-3}$), representing a 35.0% improvement. The most pronounced gains occur when scaling toward the largest model configuration, suggesting that enhanced spectral representation capacity enables more accurate modeling of global physical structures. CNN shows the strongest dependence on model capacity. At 0.01M parameters, CNN records the highest interpolation errors, with L1 and L2 reaching ($3.035\times10^{-3}$) and ($3.245\times10^{-3}$), respectively. Increasing parameter size to 1M reduces these values to ($1.928\times10^{-3}$) and ($2.418\times10^{-3}$), corresponding to improvements of 26.9% and 25.5%, respectively. Ultimately, CNN approaches Transformer performance at large scales despite its weaker small-scale performance.

However, extrapolation results under parameter scaling (Fig. 13(d)) reveal markedly different trends. FNO achieves the strongest extrapolation performance at small and medium scales but experiences substantial degradation at larger model sizes. Specifically, the L1 error increases from 0.0391 at 0.01M parameters to 0.0763 at 1M, while the L2 error rises from 0.0384 to 0.0705, representing a 80% increase. This dramatic deterioration suggests that excessive capacity may encourage overfitting to spectral characteristics of the training distribution and weaken robustness to unseen plasma configurations. CNN exhibits more stable scalability behavior. Although the extrapolation error remains relatively high at small parameter scales (L1 = 0.0431, L2 = 0.0422), increasing capacity causes only moderate degradation, reaching 0.0652 and 0.0624 at 1M parameters. Compared with FNO, CNN maintains stronger robustness under aggressive scaling. The Transformer displays the most stable extrapolation characteristics across all parameter configurations. The L1 error changes only from 0.0575 to 0.0668, while the L2 error varies from 0.0531 to 0.0617 across a 100× increase in parameter size. Although Transformer performs slightly worse than FNO at small scales, it avoids the severe degradation observed in FNO and maintains relatively consistent behavior.

Taken together, the scaling results reveal distinct trade-offs among training-data volume, model capacity, IID accuracy, and OOD robustness. The Transformer achieves the highest IID accuracy and relatively stable performance across the evaluated model sizes, but increasing the training-data volume does not consistently improve its OOD performance. The FNO exhibits strong OOD performance at small and moderate model sizes, although excessive model capacity leads to performance degradation. The CNN benefits most consistently from increases in both dataset size and model capacity and maintains comparatively stable OOD behavior. These results indicate that larger datasets and models improve IID accuracy but do not necessarily guarantee better OOD generalization.

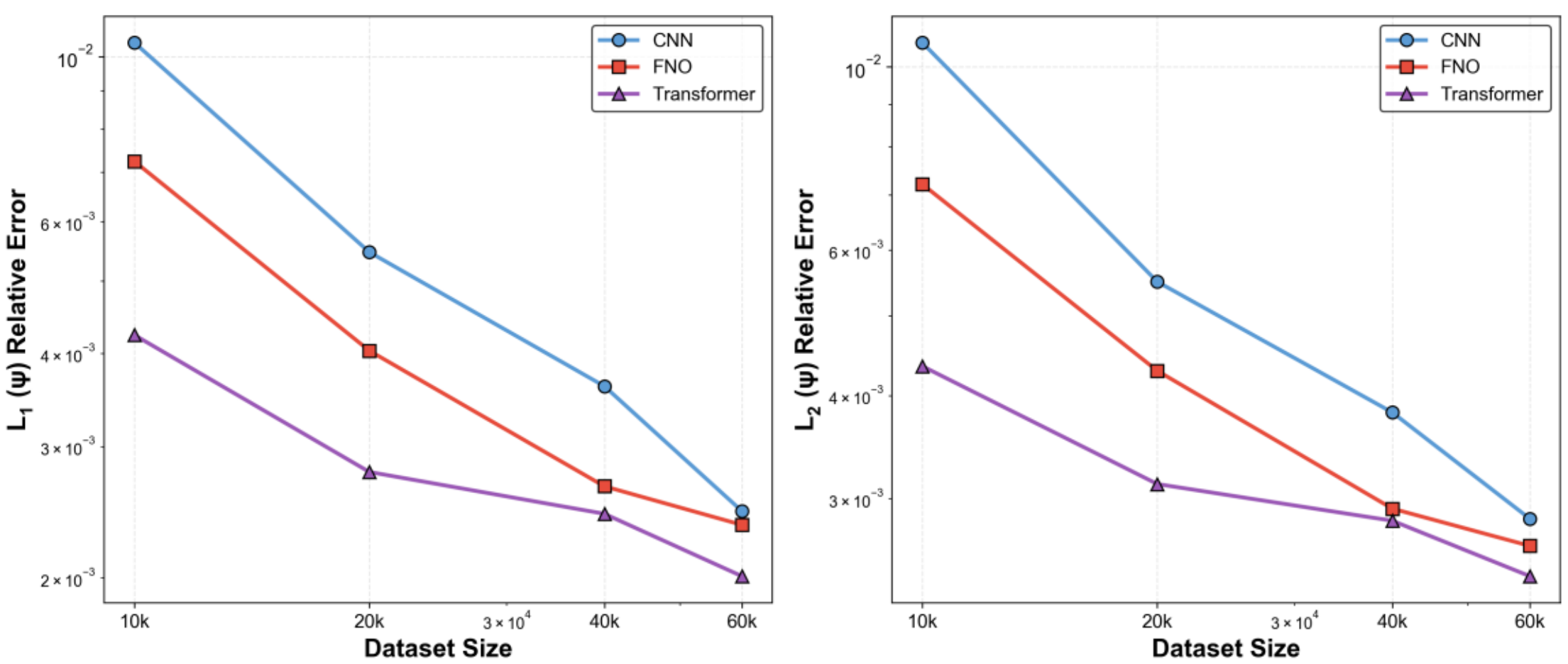

(a) Test results under varying training dataset sizes (interpolation)

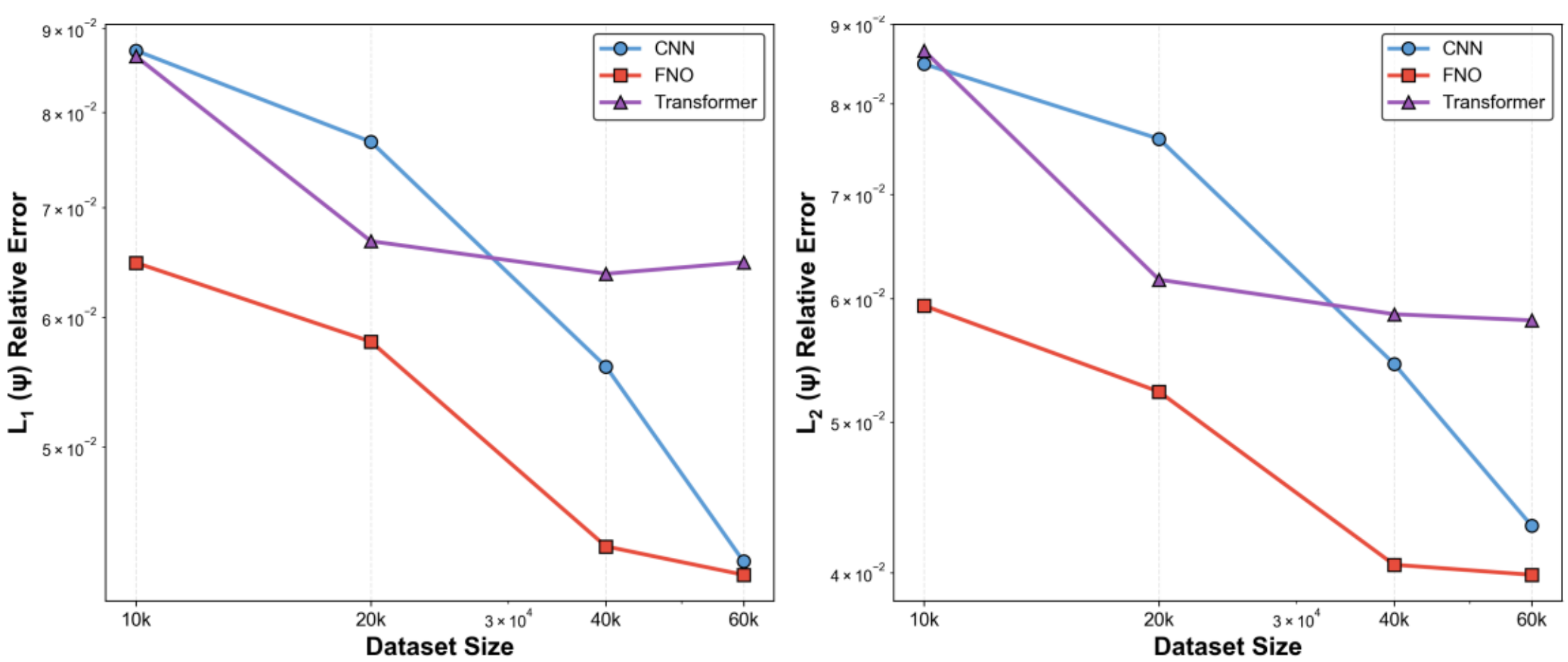

(b) Test results under varying training dataset sizes (extrapolation)

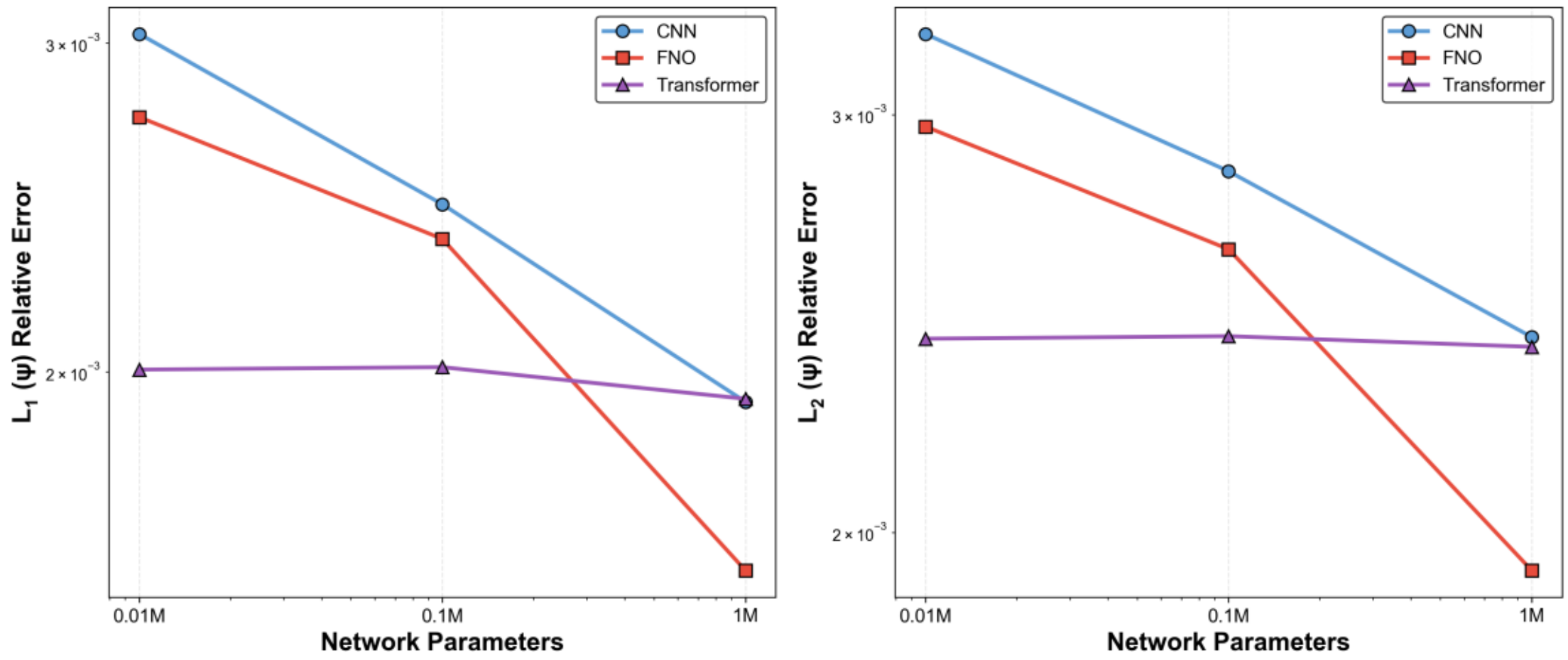

(c) Test results under varying model parameter sizes (interpolation)

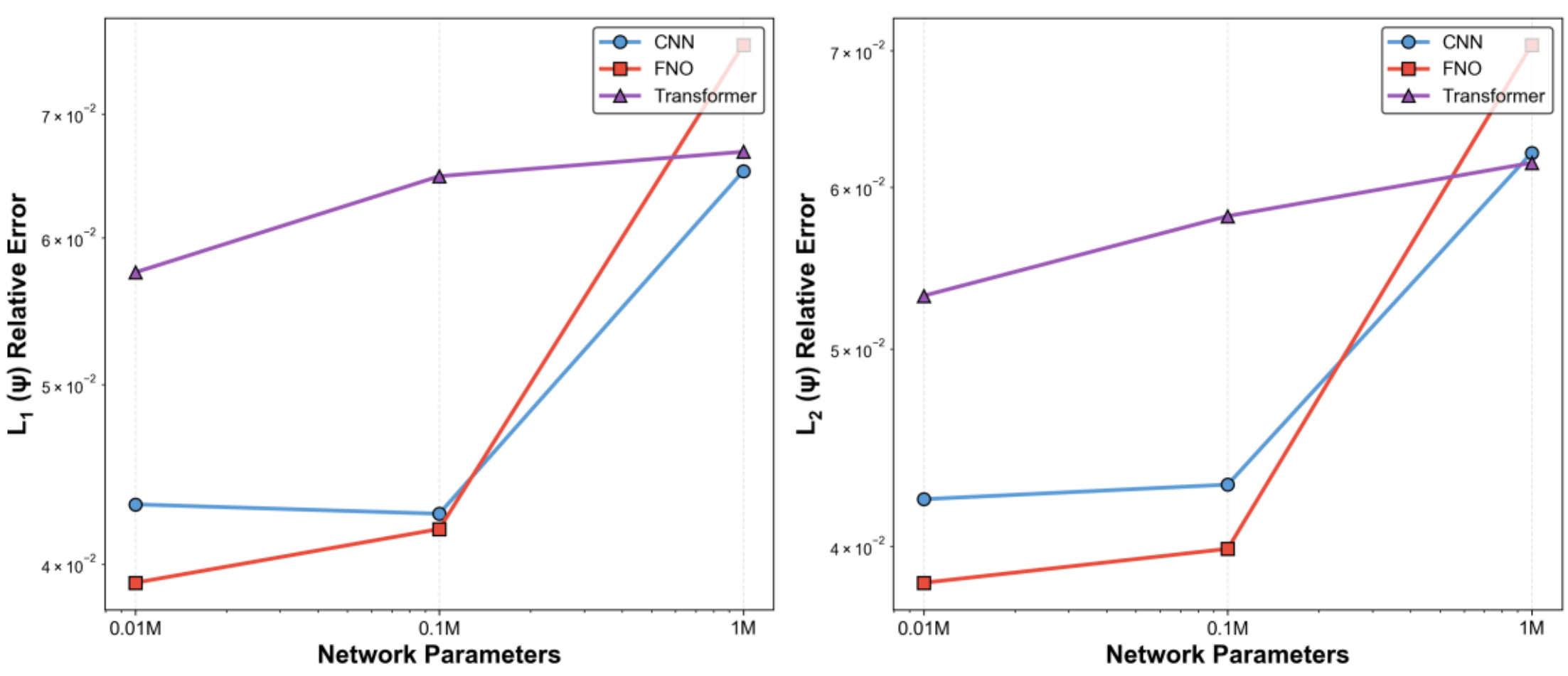

(d) Test results under varying model parameter sizes (extrapolation)

Fig. 13. Model Performance across Different Dataset Sizes and Model Parameters.

### 4.4 Computational Efficiency and Real-Time Deployment Potential

All benchmarking experiments are conducted on a workstation equipped with an Intel Xeon Gold 6530 CPU and an NVIDIA RTX 4090 GPU. Owing to the inherently serial nature of iterative solvers, the MATLAB-based FDM-SOR baseline is restricted to CPU execution. In contrast, the neural-network models are evaluated across both CPU (PyTorch) and GPU backends, including PyTorch, TorchScript, ONNX, and TensorRT. As summarized in Table 5, the FDM-SOR solver requires 10.2 s for a single inference on the CPU. By comparison, the lightweight architectures, MLP, KAN, CNN and Transformer, achieve exceptional computational efficiency, with TensorRT-optimized inference times of 0.2 ms, 0.4 ms, 0.7 ms and 0.6ms, respectively. The FNO model is comparatively slower, requiring 3.5 ms.

This represents a dramatic speed advantage over traditional iterative solvers. Notably, the inference latencies of MLP, KAN, CNN, and Transformer fall well within the millisecond-level response requirement of tokamak real-time control loops, demonstrating strong potential for deployment as real-time diagnostic and control tools.

Table 5. Inference time of various models under different inference frameworks

| Model | PyTorch (CPU) | PyTorch (GPU) | TorchScript (GPU) | ONNX (GPU) | TensorRT (GPU) |
|---|---|---|---|---|---|
| FDM-SOR | - | - | - | - | - |
| MLP | 2.8 ms | 0.3 ms | 0.2 ms | 0.5 ms | 0.2 ms |
| KAN | 795 ms | 2.0 ms | 0.6 ms | 8.7 ms | 0.4 ms |
| FNO | 642.1 ms | 4.4 ms | 3.5 ms | - | - |
| CNN | 26.8 ms | 2.9 ms | 1.8 ms | 2.2 ms | 0.7 ms |
| Transformer | 3947.3 ms | 6.2 ms | 3.1 ms | 6.36 ms | 0.6 ms |

### 4.5 Device-Relevant Validation on the EXL-50U Tokamak

To validate the practical applicability and accuracy of the developed data-driven surrogate model, we conducted comprehensive tests on the EXL-50U tokamak facility. EXL-50U is a compact spherical tokamak located at ENN Energy Research Institute, designed for fundamental plasma physics research and fusion technology development. The experimental validation employed equilibrium prediction data from actual discharge scenarios. For comparison purposes, we utilized outputs from the standard Shape Editor (SE) solver, which serves as the conventional method for waveform design in the EXL-50U control system.

For experimental validation, representative discharges #6460 and #7533 are selected to cover a typical plasma operating regime. Since the preceding offline evaluation yields consistent conclusions and similar spatial error patterns across different network architectures, only a single representative model is considered here to avoid redundant presentation and to emphasize device-level applicability. Specifically, only the CNN-based surrogate, which exhibits the most stable overall performance in offline tests, is subjected to on-device validation. The actual GS numerical solution is adopted as the reference ground truth, while the conventional SE solver used in the EXL-50U control workflow is included as the operational baseline.

Under this setup, Fig. 14 summarizes the equilibrium prediction results for discharges #6460 and #7533, including the corresponding relative error distributions. In each subfigure, the left panel compares the poloidal-flux contours obtained by two methods, and the right panel shows the relative error with respect to the GS reference solution. Fig. 14(a) and 14(b) correspond to discharges #6460 and #7533, respectively, and compare the GS solution with the prediction produced by the SE solver. The GS and SE results exhibit consistent flux-surface topology. The main flux-surface shape, magnetic-axis location, and plasma boundary are in overall agreement. The relative discrepancy is predominantly on the order of $10^{-3}$ over the plasma region, with localized lower-error bands reaching $10^{-4}$, most visibly on the outboard side, without altering the global flux-surface structure. This agreement indicates that the GS full solution is consistent with the operational SE baseline, supporting its use as the reference ground truth for assessing data-driven predictions.

On this basis, Fig. 14(c) and 14(d) correspond to discharges #6460 and #7533, respectively, and compare the GS solution with the proposed CNN surrogate. The CNN accurately reproduces the flux-surface geometry and nested topology, with the magnetic-axis position and contour shape closely matching the GS solution. The relative error predominantly falls in the $10^{-2}$~$10^{-3}$ range over the plasma cross-section. In summary, the CNN surrogate remains consistent with the GS reference solution on device-level data, with errors largely confined to the plasma edge, indicating its practical applicability.

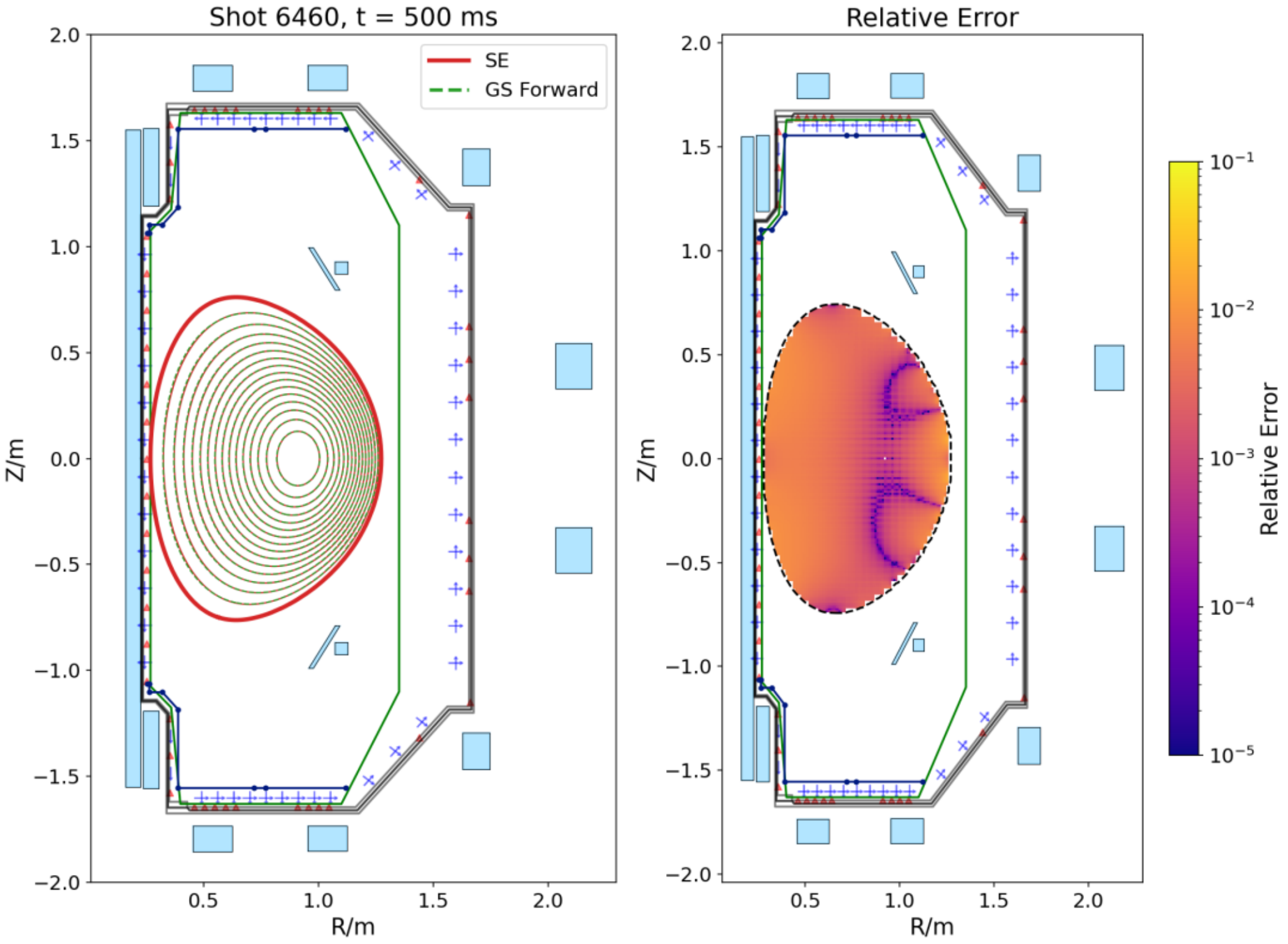

(a) #6460 GS vs. SE Comparison

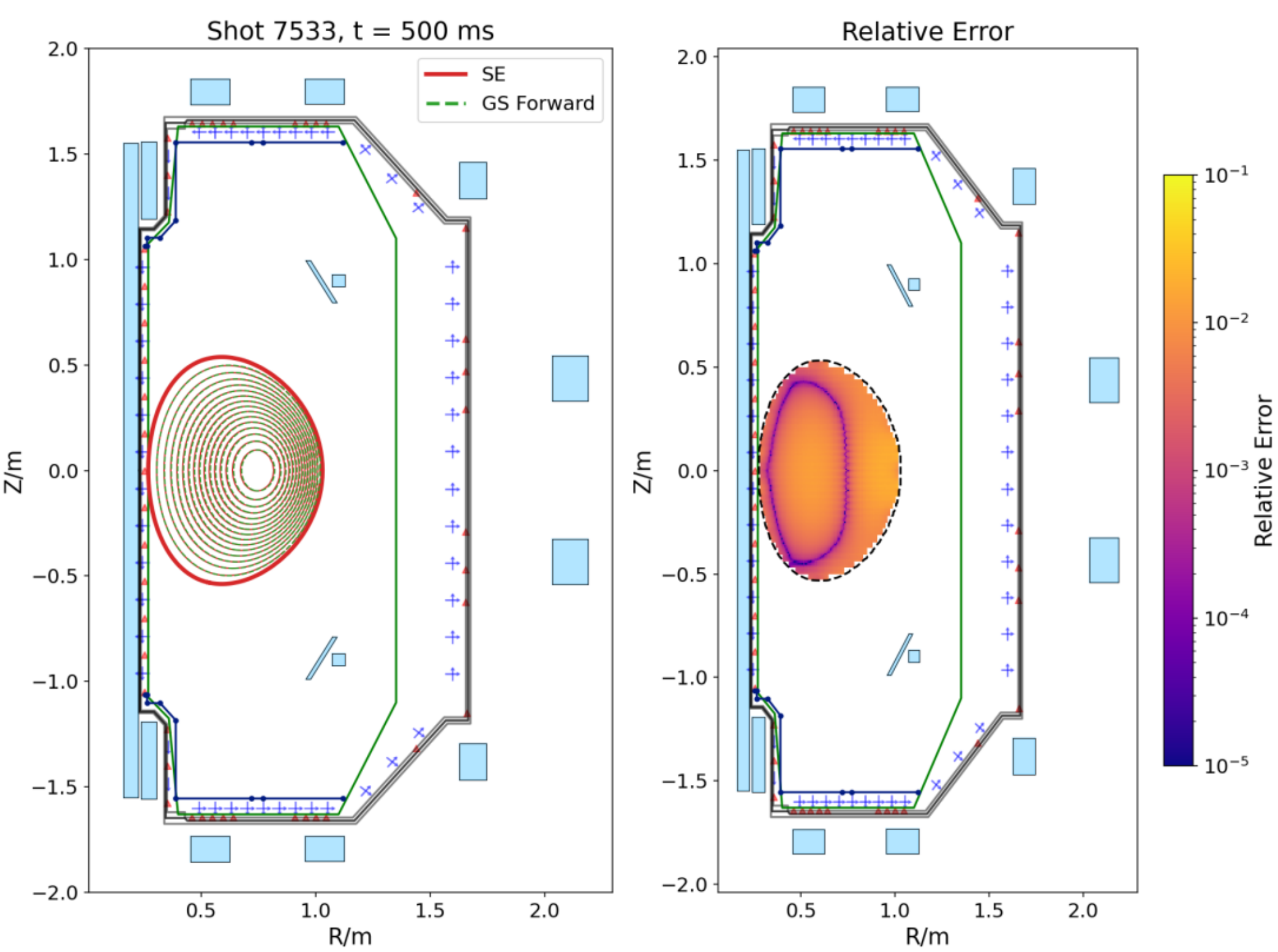

(b) #7533 GS vs. SE Comparison

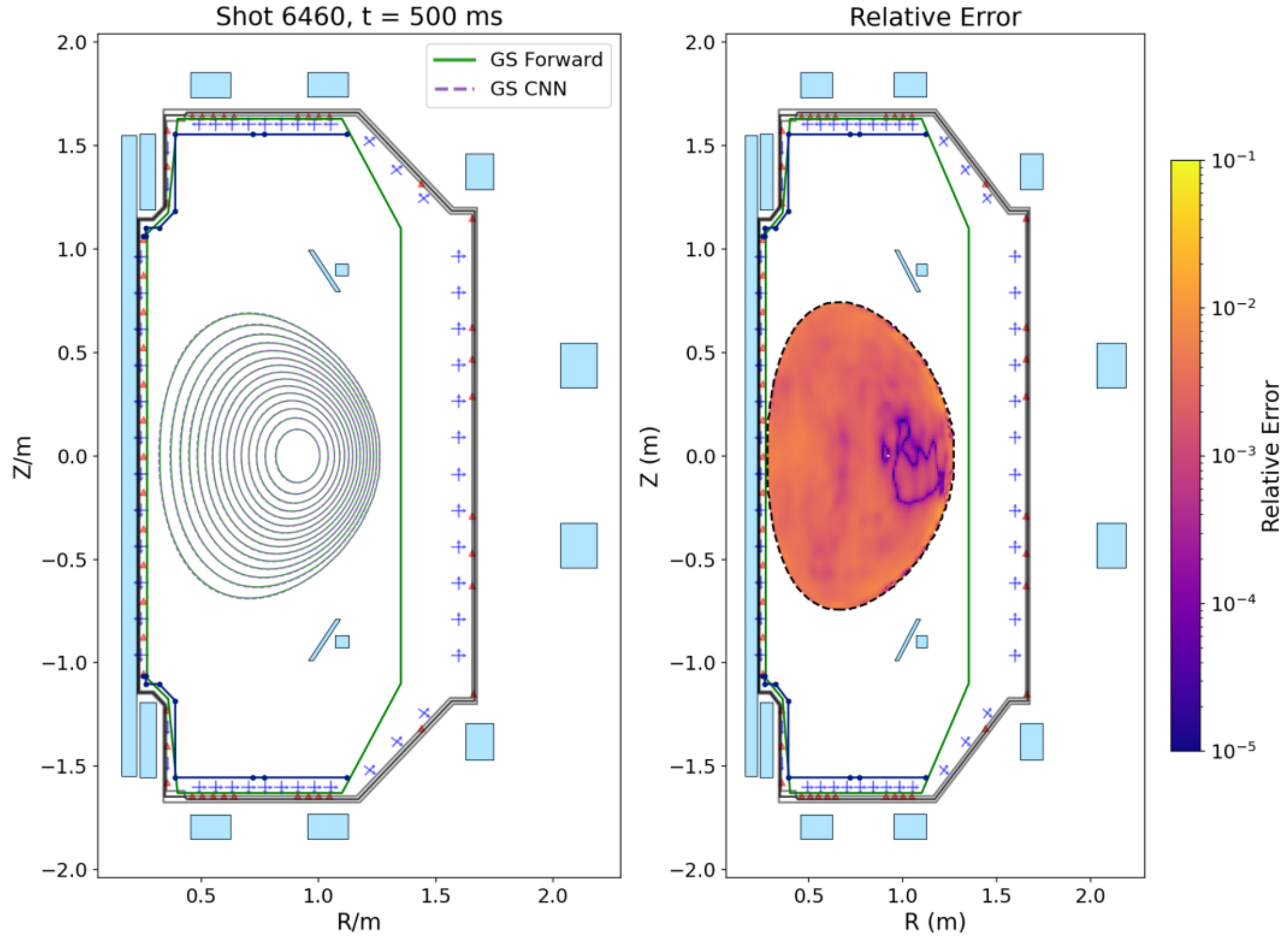

(c) #6460 CNN Predictions

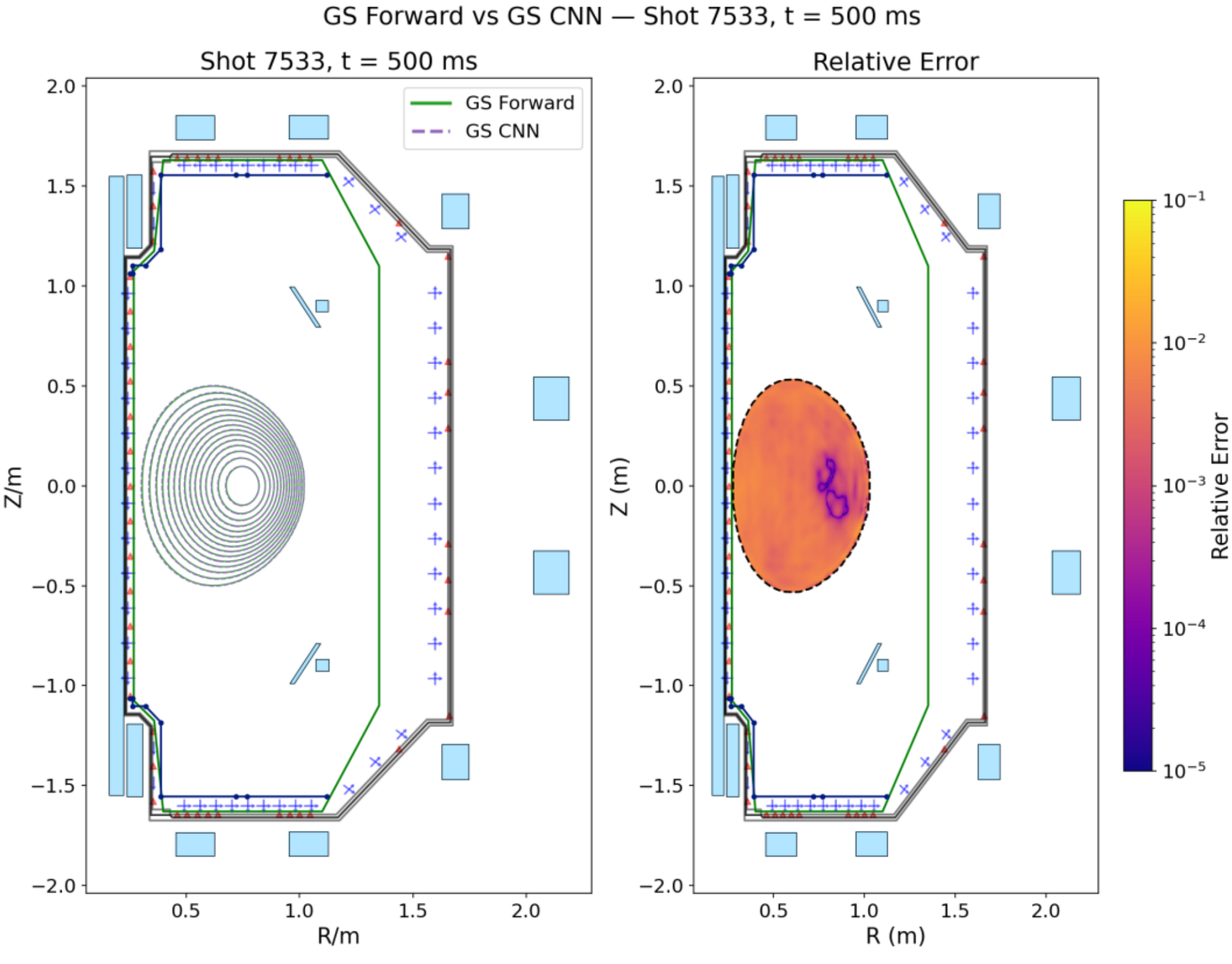

(d) # 7533 CNN Predictions

Fig. 14. Experimental Validation on EXL-50U.

This section demonstrates the technical feasibility of achieving computational acceleration while maintaining physical rigor, from theoretical GS equations through data-driven learning to actual device validation. Through dual validation (CNN precisely reproduces GS and GS corresponds well with SE operational benchmarks), we confirm that the data-driven model not only mathematically learns the GS equation solution operator with precision, but more importantly, its predictions maintain good physical consistency with actual device operational standards, validating the correctness and reliability of this approach in engineering applications.

## 5. Conclusion

This study presents a systematic benchmark of five representative deep learning architectures (MLP, CNN, FNO, KAN, and Transformer) for fixed-boundary Grad–Shafranov equilibrium prediction. A unified AI surrogate modeling framework is established to integrate high-fidelity dataset generation, model training, architecture benchmarking, scaling analysis, computational-efficiency assessment, and device-relevant validation. The principal contributions and findings are summarized as follows.

(1) A systematic AI surrogate framework with comprehensive architecture benchmarking is established for real-time tokamak equilibrium prediction. A high-fidelity database containing 100,000 independent and identically distributed (IID) samples and 10,000 out-of-distribution (OOD) samples is constructed. Under a unified evaluation protocol, the five architectures are compared in terms of accuracy, efficiency, and robustness. The Transformer achieves the highest IID accuracy, with relative $L_1$ and $L_2$ errors of 0.201% and 0.242%, respectively. The CNN provides a favorable balance among accuracy, OOD robustness, and inference efficiency, achieving a TensorRT latency of 0.7 ms

(2) A simulation-to-device validation pathway is demonstrated for AI-based equilibrium prediction. Numerical GS solutions, AI surrogate predictions, and operational Shape Editor reference equilibria are compared using representative EXL-50U configurations. The discrepancy between the GS solutions and Shape Editor references remains predominantly on the order of $10^{-3}$, while the CNN predictions exhibit relative discrepancies of $10^{-3}$–$10^{-2}$ with respect to the GS references. The predicted magnetic-axis positions, plasma boundaries, and flux-surface structures agree well with the reference equilibria, supporting the device relevance of the proposed framework

(3) The scaling behavior and generalization characteristics of AI surrogate models are investigated. OOD tests involving unseen plasma geometries, current conditions, and source-profile parameters reveal substantial architecture-dependent differences.

The CNN and FNO exhibit the strongest OOD robustness, whereas the KAN undergoes severe degradation despite its competitive IID accuracy. Scaling experiments further show that increasing dataset size and model capacity generally improves IID accuracy but does not necessarily enhance OOD robustness, revealing a trade-off among accuracy, generalization, model capacity, and efficiency.

Overall, this work establishes a systematic evaluation methodology combining cross-architecture benchmarking, IID/OOD assessment, scaling analysis, computational-efficiency evaluation, and device-relevant validation. The results demonstrate the potential of AI surrogate models for fast tokamak equilibrium prediction and provide practical guidance for future real-time plasma-control applications.